\documentclass[table, screen, nonacm, acmsmall]{acmart}

\AtBeginDocument{%
  }

\setcopyright{rightsretained}
\acmJournal{PACMHCI}
\acmYear{2025} \acmVolume{9} \acmNumber{2} \acmArticle{CSCW097} \acmMonth{4} \acmPrice{}\acmDOI{10.1145/3710995}

\acmBooktitle{Proc. ACM Hum.-Comput. Interact., V9, N2, Article CSCW097 (CSCW'25), April 2025, Bergen, Norway}
\acmISBN{2573-0142/2025/4 - CSCW097}

\usepackage{tabularx}
\usepackage{booktabs}
\usepackage{graphicx}
\usepackage{hyperref}
\usepackage{hhline}
\usepackage{svg}
\usepackage{geometry}

\usepackage{tabularray}

\usepackage{subcaption}
\usepackage{enumitem}

\usepackage{float} 
\usepackage{soul,color}
\usepackage[utf8]{inputenc}
\usepackage{chngcntr}
\usepackage{xcolor}
\definecolor{editCol}{rgb}{0.0, 0.0, 0.0}
\newcommand{\edit}[1]{{\textcolor{editCol}{#1}}}
\definecolor{redCol}{rgb}{1.1, 0.0, 0.0}

\usepackage{pbox}
\usepackage{multirow}
\usepackage{array}
\usepackage{makecell}
\usepackage{afterpage} 
\usepackage[title]{appendix}
\usepackage{subcaption,graphicx}
\usepackage{ragged2e}

\newcolumntype{L}[1]{>{\raggedright\let\newline\\\arraybackslash\hspace{0pt}}m{#1}}
\newcolumntype{C}[1]{>{\centering\let\newline\\\arraybackslash\hspace{0pt}}m{#1}}
\newcolumntype{R}[1]{>{\raggedleft\let\newline\\\arraybackslash\hspace{0pt}}m{#1}}

\begin{document}

\title[Adverse Childhood Experiences of CHINS Youth]{\edit{How the Internet Facilitates Adverse Childhood Experiences for Youth Who Self-Identify as in Need of Services}}

\author{Ozioma C. Oguine}
\orcid{https://orcid.org/0000-0003-2434-1400}
\affiliation{%
\department{Computer Science and Engineering}
  \institution{University of Notre Dame}
  \city{Notre Dame}
  \state{Indiana}
  \country{USA}}
\email{ooguine@nd.edu}

\author{Jinkyung Katie Park}
\email{jinkyup@clemson.edu}
\orcid{0000-0002-0804-832X}
\affiliation{
  \institution{Clemson University}
  \city{Clemson}
  \state{South Carolina}
  \country{USA}
}

\author{Mamtaj Akter}
\email{mamtaj.akter@nyit.edu}
\orcid{0000-0002-5692-9252}
\affiliation{
  \institution{New York Institute of Technology}
  \city{New York}
  \state{New York}
  \country{USA}
}

\author{Johanna Olesk}
\orcid{0009-0003-1707-0750}
\affiliation{%
\department{Computer Science and Engineering}
  \institution{University of Notre Dame}
  \city{Notre Dame}
  \state{Indiana}
  \country{USA}}
\email{jolesk@nd.edu}

\author{Abdulmalik Alluhidan}
\orcid{0000-0001-8948-3923}
\affiliation{%
\department{Computer Science}
  \institution{Vanderbilt University}
  \city{Nashville}
  \state{Tennessee}
  \country{USA}}
\email{abdulmalik.s.alluhidan@vanderbilt.edu}

\author{Pamela Wisniewski}
\email{pamela.wisniewski@vanderbilt.edu}
\orcid{0000-0002-6223-1029}
\affiliation{
  \institution{Vanderbilt University}
  \city{Nashville}
  \state{Tennessee}
  \country{USA}
}

\author{Karla Badillo-Urquiola}
\orcid{0000-0002-1165-3619}
\affiliation{%
\department{Computer Science and Engineering}
  \institution{University of Notre Dame}
  \city{Notre Dame}
  \state{Indiana}
  \country{USA}}
\email{kbadill3@nd.edu}

\renewcommand{\shortauthors}{Ozioma C. Oguine et al.}

\begin{abstract}
Youth implicated in the child welfare and juvenile justice systems, as well as those with an incarcerated parent, are considered the most vulnerable Children in Need of Services (CHINS). We identified 1,160 of these at-risk youth (ages 13-17) who sought support via an online peer support platform to understand their adverse childhood experiences and explore how the internet played a role in providing an outlet for support, as well as potentially facilitating risks. We first analyzed posts from  1,160 youth who self-identified as CHINS while sharing about their adverse experiences. Then, we retrieved all 239,929 posts by these users to identify salient topics within their support-seeking posts: 1) Urges to self-harm due to social drama, 2) desire for social connection, 3) struggles with family, and 4) substance use and sexual risks. We found that the internet often helped facilitate these problems; for example, the desperation for social connection often led to meeting unsafe people online, causing additional trauma. Family members and other unsafe people used the internet to perpetrate cyberabuse, while CHINS themselves leveraged online channels to engage in illegal and risky behavior. Our study calls for tailored support systems that address the unique needs of CHINS to promote safe online spaces and foster resilience to break the cycle of adversity. Empowering CHINS requires amplifying their voices and acknowledging the challenges they face as a result of their adverse childhood experiences. 

\end{abstract}


\received{January 2024}
\received[revised]{July 2024}
\received[accepted]{October 2024}

\maketitle

\section{Introduction}

According to the annual child maltreatment report released by the United States (U.S.) Federal Children's Bureau, 3,016,000 children received child welfare services in 2021, with approximately 600,000 identified as maltreatment victims \cite{maltreat2021}. In the same year, 606,031 children were served by the foster care system due to neglect, abuse, and parent incarceration \cite{afcars2022}. The 2022 National Report on Youth and Juvenile Justice System revealed that in 2019 \cite{puzzanchera2022youth}, 696,620 arrests were made on children under the age of 18, leading to their involvement in the juvenile justice system. The U.S. Department of Justice in 2016 documented that 1,473,700 children had incarcerated parents\cite{maruschak2021parents}. The above-mentioned children (those in the child welfare system, juvenile justice system, homeless children, and those with incarcerated parents) are collectively identified as \textbf{Children in Need of Services} (CHINS). Youth in need of services often experience long-term negative outcomes \cite{lee2022effect}, shaping their life experiences and making them more vulnerable to risky people and behaviors that put them at further risk \cite{hamstra2022longitudinal, laith2022temporal}. Therefore, studying the daily struggles of youth in need of services is critical to understanding their perspectives and providing the services they need. A novel approach for doing this is by analyzing their digital trace data as they seek support online. Adolescents (ages 13–17) are more likely to use online platforms extensively for social interactions, self-expression, self-disclosure, and support-seeking compared to younger children \cite{atske2022teens}; hence, we focused on studying the online support-seeking behaviors of CHINS between 13--17 years old.

The majority of at-risk youth (i.e., CHINS) have experienced traumatic events in their early lives, such as physical and/or sexual abuse, neglect, substance abuse, and parental incarceration \cite{baglivio2014prevalence}. These events, collectively termed \textbf{Adverse Childhood Experiences} (ACEs), directly and negatively impact the well-being of CHINS, leading to both physical and mental health risks \cite{felitti1998relationship, merrin2023}. Meanwhile, the internet serves as a valuable resource for CHINS and other youth, offering access to support services they may not otherwise have access to \cite{ cole2017online, gray2005health}. Peer-to-peer support platforms and forums play a crucial role in this regard, allowing youth to freely share their lived experiences anonymously if desired, connect with peers who have similar experiences, provide mutual support, and become part of a larger supportive online community, mirroring the support they might lack offline \cite{huh2023help, hartikainen2021safe, borzekowski2001adolescent, forte2014teens, ellison2016question}. At the same time, the internet could also facilitate risks associated with at-risk youths' adverse experiences \cite{hamstra2022longitudinal}. The Computer Supported Collaborative Work (CSCW) community has a well-established record of conducting research that explores how computer-mediated communication facilitates potential harm towards both vulnerable youth \cite{rankin2021resisting, hartikainen2021safe, tao2020wang, badill2017ghosh, heidi2021razi} and youth in general~\cite{prema2022, haiyan2015, blackwell2016, alluhidan2024teen}. Yet, to date, there have not been any comprehensive studies amplifying the voices of CHINS regarding how they seek support for these experiences online and how the internet may play a role in their adverse life experiences. To address this gap, we pose three high-level research questions:

\begin{itemize}

\item \edit{\textbf{RQ1:} \textit{When at-risk youth self-identify as children in need of services (CHINS), what adverse life experiences do they share about online?}}

\item \edit{\textbf{RQ2:} \textit{When CHINS seek support online, what are the main topics they post about?}}

\item \textbf{RQ3:} \textit{Based on the \edit{topics for which CHINS seek support, what are the predominant ways in which the internet facilitated} their adverse life experiences?}

\end{itemize}

To address these research questions, we leveraged a licensed dataset from a mental health peer support platform primarily used by young adults and adolescents, with over 5 million posts. To answer \textbf{(RQ1)}, we conducted a top-down analysis of 1,663 posts that we identified were made by 1,160 CHINS (ages 13-17) who self-identified as belonging within three CHINS categories: 1) Child Welfare, 2) Juvenile Justice, and 3) Parental Imprisonment. We leveraged the criteria for identifying CHINS \cite{in2023chins, nh2021chins, fl2021chins} and their ACEs \cite{SPCC2023} to create a codebook for coding the posts and gaining deeper qualitative insights. Overall, we found the posts of these at-risk youth were easy to find, as CHINS explicitly self-identified (e.g., \textit{`I'm a foster kid...`}) when sharing about their adverse life experiences (e.g., \textit{`My dad got arrested...'}). Our structured coding process also allowed us to perform chi-square tests to find statistical differences in the proportion of CHINS and ACEs posts across the three categories of CHINS with findings revealing nuanced insights into the major themes discussed by these CHINS about their adverse childhood experiences (ACEs). To address \textbf{(RQ2)} and unveil the topics that CHINS shared online to seek support from their peers, we extracted all of the posts by these users across the entire platform based on unique user IDs. Then, we applied a Latent Dirichlet Allocation (LDA) model to these 239,929 posts, which yielded four distinct topics that hold significance in our context: "urges to self-harm due to social drama," "desire for social connection," "struggles with family issues," and "substance use and sexual risks". Finally, to answer \textbf{(RQ3)} and to shed light on the role of the internet in exacerbating ACEs among CHINS, we conducted a thematic content analysis on a subset of posts (\textit{n} = 2,000), where CHINS referred to \edit{the impact of online technologies on their} behaviors in their posts. Our findings uncovered several trends in online risk behavior that aligned with the four topics identified in RQ2. The themes identified included: 1) sharing self-harm thoughts and triggering content online; 2) seeking social connections resulting in meeting strangers online, 3) experiencing cyberabuse from family and others, and 4) engaging in illegal and risky behaviors facilitated by the internet. Overall, our study revealed that the internet provides at-risk youth a platform to seek support online regarding their traumatic life experiences, while at the same time, facilitating their ongoing trauma. As such, this work makes the following novel research contributions:

\begin{itemize}
    \item \edit{Employing a mixed-method approach (thematic analysis and topic modeling), we systematically unpacked the lived adverse experiences shared by CHINS. We discovered unique challenges among the different CHINS groups, such as youth within Child Welfare struggled more with self-harm, while those involved in juvenile justice and those with a parent in prison often referenced substance abuse by themselves and their parents.} 
    \item We moved beyond understanding the online support-seeking behavior of at-risk youth to characterizing the relationship between their ACEs and current struggles, and how that exposes them to online risks. We uncovered a concerning trend towards self-harm as a way CHINS attempted to cope with their ACES.
    \item \edit{Our study highlights the straightforward identification of CHINS and the need to educate them on privacy and security practices to mitigate vulnerabilities to online predators.} Hence, we provide researchers, technology designers, and developers evidence-based recommendations for building safer online peer-support platforms that take into account the experiences of CHINS while emphasizing teen empowerment and prioritizing their online safety.
\end{itemize}

Our research makes empirical contributions to the CSCW community and the field of adolescent online safety, focusing on online peer support for vulnerable teens who have been implicated in the child welfare and/or juvenile justice systems, as well as those who experienced the incarceration of a parent. The implications of our findings extend to assisting researchers, educators, and designers in educating and crafting online features and tools tailored to the needs of the three categories of CHINS discussed in this research based on their experiences.












\section{Background}
In this section, we synthesize literature on the intersection of Children in Need of Services (CHINS) and technology use, the role of the internet on their adverse childhood experiences, and their support-seeking behaviors online.   

\subsection{Studying Children in Need of Services in the HCI Community}
\label{2.1}
\edit{
The internet serves as a double-edged sword for youth, offering both positive connections and harmful interactions. On the positive side, research highlights its benefits in enhancing social connections and providing mental health support~\cite{andalibi2016understanding, saha2021life}. Specifically, multi-channel support systems have been shown to significantly bolster these aspects~\cite{hung2022use}. For at-risk youth, the internet is particularly valuable as it provides access to essential resources and empowers social interactions~\cite{piccolo2021chatbots, sultana2022shishushurokkha, tseng2022care}. For example, research confirms that online social therapy for youth with mental health issues is both safe and effective~\cite{wadley2013participatory}. Moreover, online communities and peer networks offer comfort and support, providing crucial emotional support, advice, and a sense of belonging~\cite{ellison2016question, ok2021defined}. This is also true when it comes to using social media as the anonymity of platforms like Reddit allows vulnerable at-risk youth to share their experiences more freely and in detail without fear of judgment~\cite{andalibi2016understanding}. 
On the negative side, the internet can also contribute to negative interactions among at-risk youth. This is particularly true for adolescents, where it can lead to technology abuse, attention deficits, emotional regulation problems, and social skills deficits~\cite{hung2022use}. Studies have found that social media platforms can become hotbeds for harassment and abuse, which are particularly damaging to vulnerable individuals~\cite{scott2023trauma}. Additionally, social media usage can lead to feelings of insufficiency and lower self-esteem due to upward social comparison, especially when users compare themselves to others who appear more successful or happier~\cite{hwnag2019social, vogel2014social}.
}

A myriad of research in CSCW has been conducted on the intersections of children, technology use, and online safety \cite{rankin2021resisting, badillo2020assets, hartikainen2021safe, agha2023strike, alluhidan2024teen}; however, less research has focused specifically on CHINS, who are arguably the most vulnerable population of youth in need of online protection \cite{badillo2017abandoned, badillo2019risk}. Although the exact definition of CHINS varies depending on the jurisdiction, in general, CHINS are children who are experiencing behavioral challenges (e.g., truancy, running away from home), struggle with mental health (e.g., self-harm, depression), family dysfunction (e.g., neglect, abuse), substance abuse, homelessness, parental issues (e.g., parental substance abuse, incarceration), or are involved with the juvenile justice system, child protective services, or foster care \cite{in2023chins, nh2021chins, fl2021chins}. By proactively identifying CHINS, they can be provided with appropriate support and services to ensure their long-term well-being and development, as well as prevent more serious problems in later life.

HCI researchers have begun to study specific groups within the CHINS population, such as homeless children \cite{woelfer2011homeless, woelfer2012homeless, hendry2017u, kuo2023understanding, Oguine_Yao_Badillo-Urquiola_2023}, foster children \cite{badillo2017understanding, badillo2017abandoned, badillo2018stakeholders, fitch2012youth, fowler2022fostering}, children with mental health issues \cite{anekar2021hci, balcombe2022human, smith2014unbounding}, and disabilities \cite{burton2011evaluation, oguine2023you} to understand their online and technology use, create better policies around them, and design technological solutions, taking advantage of the existing and emerging technology to assist those groups directly or indirectly. \edit{For instance, Woelfer et al. \cite{woelfer2012homeless} investigated the use of social network sites by homeless young people through an interview study, bringing out implications for social intervention and technical design to meet the special needs of this population. In another study, Burton et al. \cite{burton2011evaluation} studied the understanding between deaf children and their hearing caregivers, proposing a storytelling application to promote their connection.} Some research in the child welfare system has concentrated on ``street-level'' algorithmic decision-making \cite{kawakami2022improving, brown2019toward, stapleton2022imagining, saxena2023rethinking, saxena2020child}, that is "algorithmic systems that directly interact with and make decisions about people in a sociotechnical system" \cite{alkhatib2019street}. For example, in an effort to assess the risk of child maltreatment, Kawakami et al. \cite{kawakami2022care} interviewed call screeners and supervisors using the Allegheny Family Screening Tool (AFST). Their study generated redesigned concepts for algorithmic decision support tool. Saxena et al. \cite{saxena2021framework} developed a cohesive framework of algorithmic decision-making adapted for the public sector (ADMAPS) and applied it in a qualitative study to analyze the algorithms in daily use in the child welfare system in mid-Western US, proposing guidelines for the design of algorithmic decision-making tools in the child-welfare system. Beyond the CSCW literature on algorithmic decision-making in the public sector, Badillo-Urquiola et al's work has focused on the online safety of foster youth from the perspective of parents and caseworkers ~\cite{badillo2019risk, badillo2018stakeholders, badillo2017abandoned}) through interview-based inquiry.

In this study, we focus on understanding the perspective of CHINS themselves by analyzing their digital trace data when seeking support via an online peer support platform. By taking this unobtrusive approach, we were able to understand how at-risk youth self-identify as CHINS, how they sought support for their adverse life experiences, and how, in some cases, the internet facilitated both online and offline risks that targeted their vulnerable status as CHINS.

\subsection{Understanding How Youth Seek Support Online}
\edit{
}
\edit
{Youth consistently use the internet in their daily activities \cite{anderson2023teens}, a trend that has also expanded to include their support-seeking behaviors \cite{pretorius2020searching, pretorius2019young}. Prior literature suggests that social support can be accessed online through various channels such as social media platforms like Instagram \cite{chung2017personal, andalibi2017sensitive} and Facebook \cite{frison2016exploring}, online groups \cite{barta2023similar}, and health forums \cite{cui2022ketch}. This integration of online resources into youth's everyday lives has opened new avenues for seeking support, serving as an alternative option to offline support-seeking, which is often hindered by barriers like stigma and a preference for self-reliance \cite{pretorius2020searching, radez2021children}. Within the SIGCHI community, researchers have consistently highlighted how youth leverage the internet to seek support in different contexts \cite{ali2024m, young2024role, razi2020let, forte2014teens}. For instance, Pretorius et al. \cite{pretorius2020searching} explored young people's online help-seeking practices and found that they used online resources to independently search for credible mental health information, seek empathetic and personalized support that validated their experiences, or look for immediate help through real-time interactions such as chat features or hotlines during crises. More recently, and with the advances in technology and artificial intelligence (AI), researchers have also begun exploring how youth perceive new types of online social support, like those available from AI chatbots \cite{bae2021social, meng2023mediated, koulouri2022chatbots}. An example of this was highlighted in a study by Young et al. \cite{young2024role} where they found that young people generally preferred AI-generated responses for less sensitive topics like relationships, self-expression, and physical health due to their empathetic and supportive nature. However, for sensitive topics like suicidal thoughts, responses from adult mentors are preferred as they provide a more genuine and appropriate level of support. Previous research highlights the evolving nature of online support-seeking among typical youth and the growing role of online resources in providing accessible support. However, little is known about how CHINS seek online support. This study aims to understand how CHINS seek online support when sharing their adverse life experiences.
}

\edit{
Support is an important reason why young people engage online \cite{pretorius2019young, huh2023help}.
}
Prior research suggests this is because the internet has given youth the chance to obtain social support in new ways by gathering information and discussing problems online, complementing in-person social support \cite{cole2017online, piccolo2021chatbots}. Seeking support online provides youth with several advantages including potential anonymity, peer support to connect with others going through similar experiences, means for self-expression as well as 24/7 availability, and professional help if needed~\cite{forte2014teens, ellison2016question, ok2021defined}. \edit{For example, one study \cite{hung2022use} found that peer groups, including online communities, could help redirect adolescents' time and energy toward more meaningful technology use.}
Nevertheless, the internet has also negatively impacted youth by exposing them to threats such as cyberbullying, online harassment, sexual solicitations, and exposure to explicit content~\cite{copp2021online, wisniewski2016dear, livingstone2014annual}. \edit{These threats often involve various attackers such as peers, intimate partners, family members, and strangers, who can escalate and migrate across platforms and into the physical world, causing significant emotional, relational, and sometimes physical harm \cite{freed2023understanding}.}
This negative impact of the internet is especially true when it comes to CHINS, as they are more vulnerable to online risks compared to typical youth~\cite{badillo2019risk, el2018vulnerable}. Research by Wong et al. \cite{wong2021seeks} suggests that at-risk youth are more likely to seek support from formal and informal online sources, because online platforms offer anonymity, help avoid embarrassment, and provide a bigger support community.
\edit{
Hence, they turn to them to discuss critical concerns such as family relations, physical and mental health issues, and housing and homelessness \cite{fowler2022fostering}.
}

Yet, much of the research in the broader community has primarily focused on studying typical teens' online support-seeking behaviors~\cite{hartikainen2021safe, razi2020let, cole2017online, kauer2014online}. \edit{For instance, Huh-Yoo et al. \cite{huh2023help} explored youth's private pleas for support and the responses received from peers via social media private messages. They found that conversations often began casually and evolved into support exchanges, with youth seeking help for mental health concerns, relationship issues, daily life problems, and abuse, receiving a range of support types including emotional, informational, and tangible assistance. In addition, past research has also highlighted the vulnerability of at-risk youth to online risks and confirmed that foster teens often encounter high-risk online situations, such as interactions with unsafe individuals, which can lead to severe consequences like rape and sex trafficking~\cite{badillo2019risk, badillo2020assets, caddle2023duty, yates2012child}.
}
However, there has yet to be extensive research that directly incorporates the perspectives of CHINS to understand how the internet may influence their adverse life experiences. Furthermore, little is understood about how the internet provides CHINS and other vulnerable youth with the means of seeking support~\cite{moraguez2022adolescents, karusala2021courage, wong2021seeks}. Our work is the first known study to systematically identify CHINS to examine their online support-seeking behaviors, providing the HCI community with an insight into the topic from the perspective of vulnerable youth groups.

\section{Methods}
In this section, we provide an overview of our data collection methods, including details about the dataset acquisition and the data scoping procedure. We also address ethical considerations related to data and outline the data analysis approaches we employed to address our research questions. 


\subsection{Study Overview}
\edit{We leveraged a licensed dataset sourced from an online peer support platform dedicated to mental health, with a primary focus on adolescents and young adults. We anonymized the name of this platform for the protection of our participants. The majority of users were from the United Kingdom and the U.S. (English being the primary language of communication). The platform allows users to create posts and receive responses in the form of comments while also facilitating the sharing of user-generated content and discussions on an array of topics (e.g., mental health, sexuality, religion) in a semi-anonymous way, wherein users have the option to make their usernames visible or anonymous.} The original dataset encompassed an extensive corpus of data, featuring over 5 million posts and 15 million comments made by approximately 400,000 users. 


\subsection{Ethical Considerations}
This study was deemed `exempt' as non-human subject research by the authors' Institutional Review Board (IRB) as it focuses on the analysis of a secondary dataset without personally identifiable information (e.g., usernames, real names). Nevertheless, we took extra precautions to ensure all authors completed their IRB CITI training on human subject research before accessing the dataset. Additionally, in order to safeguard the privacy of users, extensive measures were undertaken to eliminate any potentially identifying information (e.g., screen names for other platforms, hyperlinks, etc.) from the posts used as exemplar quotations. Quotations featured in this research were either paraphrased or slightly altered (e.g., adding abbreviations and introducing false details that do not affect the context) to ensure that the quotes remained unlinkable to specific individuals. We also ensured that none of the quotes in this paper could be traced back to their original source through Google searches. Furthermore, due to the explicit content found in many of these posts (such as descriptions of self-harming behaviors, sexual assault, sexual acts, depression, etc.), we made sure our research team took adequate breaks when analyzing the data.

\subsection{Data Scoping Process}
Below, we describe our data scoping process to first identify children in need of services (CHINS) in the dataset and then retrieve all of their posts on the platform.

\subsubsection{Initial keyword search to identify CHINS}

We first identified search terms for data extraction based on the ten criteria used to identify CHINS (\autoref{tab:search_terms}). These criteria encompass 
involvement in the foster care/child welfare system, engagement with child protective services, experiences within juvenile detention centers, 
and interaction with the legal process (see  \autoref{tab:mapping} in the Appendix). 
\edit{For our study, we narrowed down our dataset to encompass users aged between 13 and 17 years. Subsequently, we consolidated all keywords into an SQL search query, resulting in a dataset comprised of 5,015 posts from 3,353 unique users.}

\label{3.3.1}

\begin{table}[ht]
\caption{Scoping Search Terms for RQ1 and List of Online Technology Keywords used for RQ3. Terms resulting in no records were removed. Acronym definitions are listed in \autoref{tab:acronyms} in Appendix A.}
\label{tab:search_terms}
\centering
\footnotesize
{\renewcommand{\arraystretch}{1.5}
\begin{tabular}{p{3cm}|p{8cm}}
\hline
\rowcolor[HTML]{EFEFEF}
\textbf{Category} & \textbf{Keywords} \\ \hline
\rowcolor[HTML]{FFFFFF}
\textbf{CHINS/ACEs Terms (RQ1)} & Foster care, foster home, foster parents, foster mom, foster dad, foster family, foster kid, foster child, adopted, child welfare, child protective services, child protection, CPS, DCF, DCS, CYF, DCFS, department of social services, DSS, department of human services, DHS, DHR, case manager, case worker, juvie, juvenile delinquency, prison, parole, truancy court, court case, misdemeanor, felony, arrested, police, emancipated. \\ \hline
\rowcolor[HTML]{EFEFEF}
\textbf{Online Keywords (RQ3)} & Facebook, Instagram, Tinder, Bumble, Grindr, Snapchat, Craigslist, Skype, Hinge, WhatsApp, Kik, Discord, Messenger, Omegle, Vimeo, Vine, Tumblr, Myspace, 4chan, Reddit, forum, blog, video chat, Facetime, message, DM, send, sent, PM, online, meet on, met on, webcam, gaming, cyber, blackmail, internet, amosc, f2f, lmirl, text, phone, FB, met online, TikTok. \\ \hline
\end{tabular}
}
\end{table}

Next, \edit{using a top-down approach, the first author systematically coded the dataset for relevancy} to identify and categorize posts that specifically referred to CHINS \textbf{(RQ1)}. Guided by previous research showing the overlap between child involvement in the criminal justice system with the child welfare system~\cite{Literatu99:online}, we classified the posts into three categories: Child Welfare, Juvenile Justice, and Parental Imprisonment. We removed duplicate posts to ensure the uniqueness of the content. Following the completion of the relevancy coding process, our dataset consisted of 1,663 posts contributed by 1,160 unique users. 
To gain more insights into the topics that at-risk youth discuss when seeking support online \textbf{(RQ2)}, we identified the unique userIDS (\textit{N} = 1,160) from the relevant posts and used them to extract all the posts made by those users on the platforms, which resulted in a final dataset comprising 239,929 posts.

\begin{figure}
    \centering
    \includegraphics[scale=0.52]{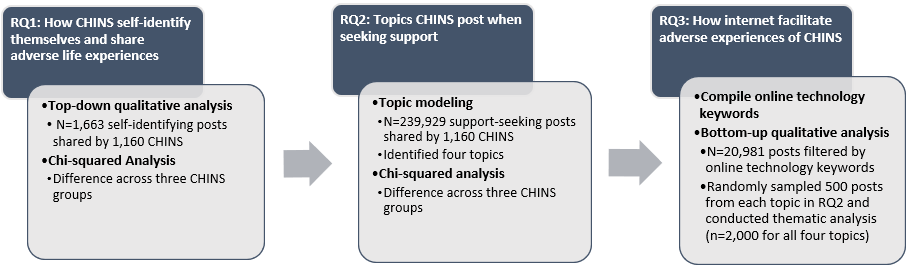}
    \caption{Overview of data scoping and analysis approaches for three RQs}
    \label{Data Analysis approaches}
\end{figure}

\edit{To understand how online technology facilitated their adverse life experiences \textbf{(RQ3)}, we compiled a list of online technology keywords to scope relevant posts. First, we identified a set of online keywords associated with prevalent social media platforms spanning the period from 2011 to 2023, such as Instagram, Kik, etc. \cite{razi2020let, 105Leadi71:online}. Next, we compiled another set of online keywords by exploring commonly used online technology jargon among adolescents \cite{razi2020let, 2023Teen6:online, 2023Teen79:online}. In addition, we added a set of online technology keywords by reviewing 1,663 posts collected to address \textbf{RQ1}. The complete list of online technology keywords is shown in \autoref{tab:search_terms}. Using these keywords, we extracted and compiled a subset of data composed of 20,981 posts from the original dataset used for \textbf{RQ3}.} We divided the posts into four topics generated by topic modeling in RQ2 and sampled 500 posts from each topic to conduct a thematic analysis. We conducted this thematic analysis of 500 posts for all four topics, utilizing 2,000 posts in total to have in-depth insights into how online technology facilitated the adverse life experiences of CHINS. The overview of the data scoping process is present in \autoref{Data Analysis approaches}. 

The age distribution of 1,160 teens in our dataset is as follows: 13 (7\%), 14 (14\%), 15 (18\%), 16 (30\%), and 17 (31\%). The self-reported gender of youth in the dataset included 72\% (\textit{n} = 829) females, 14\% (\textit{n} = 167) males, and 14\% (\textit{n} = 164) self-identifying as ``other.''

\subsection{Data Analysis Approach}


\subsubsection{ACEs and CHINS qualitative coding (RQ1)}



We conducted a deductive (top-down) qualitative analysis ~\cite{bingham2023data} on 1,663 posts based on the definitions of CHINS and ACEs (see~\autoref{tab:mapping}). 
The identity of CHINS consisted of youth implicated in the child welfare system, juvenile justice system, or who had a parent imprisoned. \edit{We then template coded our data using the American Society for the Positive Care of Children's Quiz for Adverse Childhood Experiences (ACEs) ~\cite{SPCC2023}. Specifically, we coded for how youth who are implicated in the child welfare and juvenile justice systems, as well as those with an incarcerated parent disclosed their adverse childhood experiences online such as: 1) self-harm or other imminent risks affecting their well-being; 2) self-harm by someone in their households; 3) parental or guardian neglect, inability or unwillingness to provide care; 4) physical harm from household members; 5) emotional abuse (e.g., swearing, insults, humiliation) by household members; 6) domestic violence in the household; 7) sexual abuse, trafficking, or related offenses; 8) use of illegal substances, drug abuse, or underage drinking; 9) parental or legal guardians' substance abuse; 10) loss of one or both parents; and 11) disabilities or lack of necessary medical support. We first randomly sampled 10\% of our entire dataset (\textit{n} = 164) and then the first and fourth authors coded this subset of data under the supervision of the last two advising authors who helped guide their analyses and interpretation of the results. We achieved a Cohen's kappa score of 0.87, surpassing the acceptable threshold \cite{gwet2010handbook}. The rest of the data was then coded by the first author, having the last two authors review the coding. 
This process was iterative, requiring ongoing communication and consensus among the authors.} We present our final codebook in \autoref{tab:codebook}. 

We also conducted a Chi-squared between-group analysis ($\chi^2$) to examine any significant differences among the three risk categories, i.e., child welfare, juvenile justice, and parental imprisonment, based on the listed codes. The $\chi^2$ test of independence is used when we need to conduct a between-group test among two or more variables \cite{sharpe2015chi}. To demonstrate the significant differences among the three risk categories, we used the standardized residuals calculated by dividing the difference between expected and observed values by the square root of the expected value \cite{sharpe2015chi}. Through this $\chi^2$ test, we aimed to interpret the nuanced differences between different risk categories based on the youth disclosures of different adverse experiences they faced. 



\subsubsection{Topic modeling (RQ2) and qualitative analysis (RQ3)}
\begin{figure}
    \centering
    \includegraphics[scale=0.63]{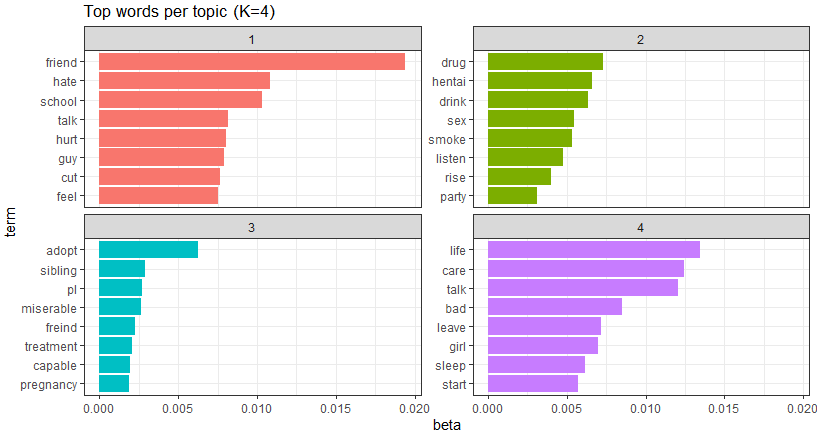}
    \caption{Top Words and the Beta Values Generated by the LDA model}
    \label{Betta_all dataset}
\end{figure}

Topics discussed by at-risk youth when seeking online support were identified using the Latent Dirichlet Allocation (LDA) model~\cite{blei2003latent}. We applied the LDA topic model as it combines an inductive approach with quantitative computations, hence, making it suitable for exploratory analysis of large-size textual data \cite{maier2018applying}. 
Before running the LDA model, the texts from all 239,929 posts were pre-processed (e.g., removing the stop words and infrequent words, lemmatizing, tokenizing, etc.) and a document term matrix was created. 
After creating the document-term matrix, the number of topics the LDA model should classify (\textit{K}) was specified by looking at the perplexity score (how surprised the model is at seeing the data, smaller the better) \cite{blei2003latent} and the coherence score (how often the topic words for each topic appear together in a document, closer to zero is better) \cite{mimno2011optimizing}. The two scores confirmed that the LDA model with the four topics (\textit{K} = 4) was a good fit. Next, we calculated the beta values i.e., the probabilities that the terms would be generated for the topic~\cite{silge2017text} and for the top words generated of the four topics. The higher the beta, the more often the word would appear within the topic. We reviewed the top words and their beta values and re-visited the posts in all three CHINS categories to deductively name the four topics generated by the LDA model. The top words and their beta values are presented in \autoref{Betta_all dataset}.

After iteratively reviewing the topic words and posts, four topics were named: "Urges to self-Harm due to social drama," "Substance use and sexual risks," "Struggles with family issues," and "Desire for social connection." \autoref{tab:fourtopics} in Appendix D shows the top four topics generated by the topic modeling and some example quotes that represent each topic. Finally, gamma values \cite{silge2017text} were calculated to see how prevalent the four topics were within each CHINS group. The higher the gamma, the more likely the group is to talk about the topic. The gamma values and the approximate number of posts that discuss each topic with three CHINS groups are presented in Appendix D (\autoref{table_gamma}).

\edit{To answer \textbf{RQ3}, we conducted a thematic analysis \cite{braun&clarke} on a subset of 10\% (\textit{n} = 2,000) of posts sampled from the dataset (\textit{n} = 20,981), where youth referred to online technology when sharing their adverse experiences. We began by mapping the keywords/themes from RQ1 to the beta values generated by the topic modeling in RQ2 (\autoref{Betta_all dataset}). This allowed us to divide the 20,981 posts into four distinct topics. We then randomly sampled 500 posts from each of the four topics and conducted a thematic analysis for each to identify how the internet facilitated harm. Specifically, we aimed to pinpoint the primary way online technology contributed to adverse life experiences for CHINS, aligned with the topics identified in RQ2. The first author coded the posts, and together with the co-authors, reached a consensus on the emergent codes and interpretation of themes. This process of data sampling and thematic analysis was repeated for all four topics. Through this approach, we identified four emergent themes that each aligned one-to-one with the four topics from RQ2: }"Sharing of self-harm and triggering content," "Engaging in illegal and risky behaviors,"  "Experiencing cyberabuse from family and others," and "Seeking social connections resulting in meeting strangers online" describing instances where online technology facilitated adverse life experiences of CHINS. The mapping of four technology-related risks to four topics generated from topic modeling is presented in Appendix E (\autoref{tab:topics}).

\section{Results}
In this section, we begin by describing \edit{how CHINS identify themselves while sharing their }adverse childhood experiences online \textbf{(RQ1)}. Then, we explore the topics posted by CHINS when seeking support online \textbf{(RQ2)}. Finally, we discuss the role of technology in exacerbating adverse experiences of CHINS \textbf{(RQ3)}.

\begin{figure}[!ht]
    \centering
    \includegraphics[scale=0.26]{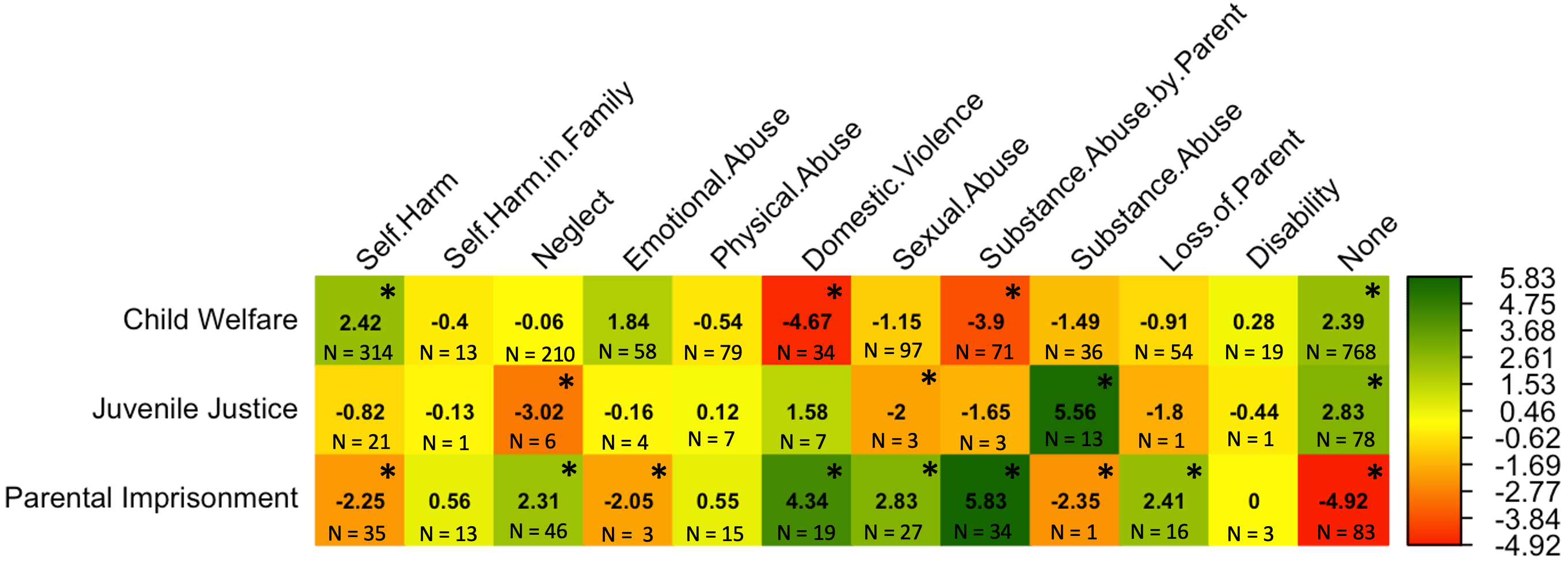}
    \caption{Results (standardized residuals) of the between-group analysis of teens' identities based on the adverse experiences teens discussed ($N = 1,663$). (*) indicates significant association. Note that green denotes a positive association, while red denotes a negative one.}
    \label{fig:rq1_chisquared}
\end{figure}
\subsection{The Adverse Childhood Experiences of Youth when they Identify Themselves as CHINS (RQ1)}
This section outlines the different types of adverse life experiences shared by at-risk youth who also identified themselves as CHINS. \edit{First, we examine how these youth self-identified as CHINS. Most youth involved with child welfare systems (84\%, n = 1390 posts) mentioned their status as adopted or foster children, using languages such as foster care, foster home, foster parents, and foster family. Additionally, they often disclosed their involvement with agencies such as Child Protective Services (CPS), the Department of Children and Family Services (DCF), the Office of Children, Youth, and Families (CYF), the Department of Social Services (DSS), or the Department of Human Rights (DHR). For instance, to share their involvement with child welfare services, one youth mentioned: }
\edit{\begin{quote}
\textit{``That's it I'm done !!! My mom is taking me to a \textbf{foster home} tomorrow..... What's the point if I'm just ending my life tonight ?''} - 14-year-old female, Child Welfare
\end{quote}}
\edit{Conversely, many youths also disclosed that their parents or legal guardians were incarcerated. In these posts (12\%, n = 200), they frequently mentioned their parents or legal guardians being arrested, charged with felonies, on parole or probation, in prison, or involved in court cases. The majority of these posts indicated that one or both parents were in prison for extended periods, while some posts described parents being frequently arrested, attending court cases, or recently entering prison. For instance, one youth disclosed about her parent's being in prison as follows:}
\edit{\begin{quote}
\textit{``\textbf{My dad is in prison} now for something he did to me. But its his on and off thing. My mom has been in hospital forever. I have a sudden feeling that I lose everyone and everything around me...''} - 13-year-old female, Parental Imprisonment
\end{quote}}

\edit{For the youth involved in the juvenile justice system (7\%, n = 114 posts)  used similar to the above imprisonment references for themselves. They also referred to juvenile issues such as juvenile delinquency, truancy courts and more.  In the following quotation, a youth shared their involvement with juvenile justice disclosing an arrest: }
    \edit{\begin{quote}
\textit{``\textbf{I got arrested for stealing today} and I've written my note I want to die. 6 months clean from self harm and now my arms slit open.''} - 17-year-old female, Juvenile Justice
\end{quote}}




\edit{Next, we present different adverse childhood experiences the at-risk youth shared on the posts when they also identified themselves as CHINS. In these posts, it becomes apparent how the overlap between their identity as CHINS is often irrevocably iterwined with their adverse life experiences (ACEs).} As shown in the \autoref{fig:rq1_chisquared}, $\chi^2$ test indicated significant differences between the negative experiences youth mentioned based on the relative expected proportions of the total number of posts ($\chi^{2} (df=22, N=2,183) =146.1, \textit{p} < 0.001$), where green cells denote a higher proportion of adverse life experiences, red denote a lower proportion, and an asterisk (*) denotes statistical significance. Below, we highlight the significant differences of the adverse experiences disclosed by the youth when they also explained their association with child welfare, juvenile justice, and parental imprisonment.

\subsubsection{Self-harm (self and family)}
The youth shared various unpleasant experiences from their daily lives, with self-harm being the most commonly discussed topic. Around a quarter of the posts (22\%, \textit{n} = 370) detailed overwhelming emotional struggles leading to \textbf{physical self-harm}. Additionally, 1\% of the posts (\textit{n} = 17) addressed instances of one or more of their \textbf{family members' urge to self-harm}. As illustrated in \autoref{fig:rq1_chisquared}, the relative proportion of posts related to the youth involved with child welfare (19\%, \textit{n} = 314) mentioning self-harm experiences reached positive significance, indicating that among the three types of CHINS, youth involved with child welfare posted about self-harm proportionally more often, compared to youth implicated in juvenile justice or who had a parent in prison. Below, we present a quotation illustrative of the larger dataset to show the way in which youth self-identified as CHINS and disclosed about self-harm when they sought support on the platform.

\begin{quote}
    \textit{`` I'm adopted. \textbf{I'm a severe self harmer.} nothing seems to go right in my life, and I'm only 13. I don't want to live, I have severe depression. I feel like I'm crazy but I can't help it.''} - 14-year-old female, Child Welfare
\end{quote}

Youth within the child welfare system frequently detailed the traumatic experiences they endured within their families, primarily involving their biological parents or step-parents. These distressing encounters often served as significant stressors leading to their engagement in self-harm behaviors, ultimately resulting in their involvement with the child welfare system. These narratives shed light on the complex and challenging circumstances these young individuals faced within their family dynamics, highlighting the critical role these experiences played in their development and current struggles.
\begin{quote}
\textit{``When I was five I was molested by my sisters dad. My mom was a druggy and she mentally, verbally and physically abused my little sister and I. Then at the age of 7 I was raped and molested and forced to do disgusting things till the age of 8 by two different people. Then DCF took me away to live with my grandmother. \textbf{I became very depressed and started cutting and starving myself like eating 2 meals a week.} 
''} - 16-year-old female, Child Welfare
\end{quote} 
However, there were also many posts that explained how the child welfare services themselves caused more anxiety and apprehension to youth as they were often asked to switch to other schools, move out from their parent's places, or even relocate to other cities/states. Youth often mentioned these sudden changes as stressors for their urge to cut, burn, or strangle themselves. 

\subsubsection{Neglect and abuse}
Some of the posts (36\%, \textit{n} = 615) included youth describing instances of neglect and abusive behaviors by their family members, often involving acts of violence. Among these posts, youth mostly wrote about how they felt \textbf{neglected} (16\%, \textit{n} = 262) by their biological parents, foster families, or both. 
When comparing the proportions of youth's neglect disclosures for each CHINS type, youth who were implicated in child welfare (13\%, \textit{n} = 210) had the highest relative proportion as shown in \autoref{fig:rq1_chisquared}. 
Yet, youth with a parent in prison (3\%, \textit{n} = 46) had a statistically significant association with these disclosures based on the standardized residuals.  
This suggests that when youth shared about their parents being in prison, the conversations were most likely to revolve around the youths' emotions of feeling unloved by their family.
Within this subset of posts, one of the key trends we observed was that youth were neglected by their biological or step-parents when their other parents were away in prison. 

\begin{quote}
\textit{``A week after my dad was arrested, we somehow got bed bugs in my mom's room... \textbf{So my mom got rid of her bed and decided to take over my room and said "just sleeping in there" and I was sleeping on her bedroom floor.} She was in there for a month and a half then she decided to take MY bed out of MY room and put it in her room.''} - 15-year-old female, Parental Imprisonment
\end{quote}

Youth also shared some posts (4\%, \textit{n} = 65) where they mentioned incidents of \textbf{emotional abuse} that took place in their families. In the majority of these posts, teens cited that they were implicated in child welfare, and their foster parents mistreated them by yelling, scolding, silent treatment, and/or hurtful comments (3\%, \textit{n} = 58). These youth often found themselves not fitting in their foster families, mostly because of their failed attempts of living with their biological parents, or because of their behavioral disorders.
Similarly, youth also often posted about being \textbf{abused physically} (6\%, \textit{n}=101), where the majority of these posts cited youth being implicated in child welfare (5\%, \textit{n} = 79). Within these posts, youth mostly shared about the abusive behavior of their biological parents, and therefore, CPS relocated them to foster homes. However, they also often wrote stories about their foster parents hurting them physically. To this end, youth mostly talked about incidences where their biological/foster parents threw sharp objects at them, pushed/kicked them, hit them with heavy objects, burned them with hot coals or irons, and so on. 

The above forms of physical abuse were often violent in nature and even life-threatening. About 4\% of the posts (\textit{n} = 60) included descriptions of how their lives were at risk because of the aggressive and \textbf{violent behavior} of their parents. Youth who were implicated in child welfare (2\%, \textit{n} = 34) had the highest relative proportion of posts describing how domestic violence by their biological parents led them to be involved with child welfare. 
Yet, as \autoref{fig:rq1_chisquared} showed, the relative proportion of posts related to youth's parental imprisonment (1\%, \textit{n} = 19) reached positive statistical significance, suggesting that when youth posted about how they were being victims of domestic violence, these posts had the highest probability of involving youth with parents in prison. In most of these posts, youth often recount instances of physical abuse inflicted by their parents, resulting in a need to seek medical help and leading to the arrest of their parents. 
\begin{quote}
\textit{``\textbf{I'm in the hospital with broken arm leg and ribs. Also a small bleed to the brain.} Got an operation in the morning to put metal plates in leg and arm. My dad and brother were arrested.''} - 14-year-old female, Parental Imprisonment
\end{quote}

Similar to youth's propensity to post about neglect and above types of abuses, we found that youth who were involved with child welfare (6\%, \textit{n} = 97) had the highest relative proportion of posts describing \textbf{sexual abuse} as shown in \autoref{fig:rq1_chisquared}. Yet, youth's parents being in prison (2\%, \textit{n} = 27) was statistically significantly associated with these disclosures based on the standardized residuals. This indicated that youth discussed their experiences of being victims of sexual abuse more frequently when they mentioned that their parents were in prison.
One of the key themes we noticed here was that they mostly talked about being sexually assaulted by their male family members, e.g., biological/foster father, elder brother, or uncle. For instance, a youth described how their traumatic experiences of being sexually assaulted by their own father for years were still tormenting their day-to-day lives.  

\begin{quote}
\textit{``\textbf{I was sexually assaulted by my father multiple times when i was eight and younger.} He applied for parole a month ago but thankfully didn't get it. 
''} - 13-year-old transgender, Parental Imprisonment
\end{quote}

\subsubsection{Substance abuse (self and parents)}

Youth also often shared about their own or family members' addiction and overdose of alcohol and recreational drugs consumption (9\%, \textit{n} = 158). Among these posts, youth most often shared distress because of their parents' chronic alcohol/drug abuse (7\%, \textit{n} = 108). Only 3\% of the posts (\textit{n} = 50) had mentions regarding their own substance usage. By looking at \autoref{fig:rq1_chisquared}, and comparing the relative proportion of posts regarding \textbf{parents' substance use} across the three different kinds of CHINS, we found that youth who were implicated in child welfare (4\%, \textit{n} = 71) had the highest relative proportion. Yet, youth's parents being in prison (2\%, \textit{n} = 34) was statistically significantly associated with discussing their parents' substance use, suggesting that when youth discussed their parents' alcohol or drug abuse, they mentioned about their parental imprisonments more frequently.
One of the interesting trends we found in these posts was that their parents' used illegal drugs or committed crimes, often violent, while they were intoxicated, leading to their arrest or imprisonment. For instance, a youth's post described how their father committed a capital crime against her sister after he consumed alcohol and was arrested afterward. 
\begin{quote}
\textit{
``
When i was 9 my dad came home one night and He got drunk and called some friends over. 
\textbf{My dad and his friends took my sister tied her feet up. Put a gag in her mouth. Plastic bag over her head and threw her in the fireplace...} 
The cops came to our house arrested my dad and took away for good.''} - 14-year-old female, Parental Imprisonment
\end{quote}

For the posts where youth shared about their \textbf{own substance use problems} (3\%, \textit{n} = 50), we found a positive statistical significance with them being implicated in juvenile justice (1\%, \textit{n} = 13), 
indicating that when youth posted about their own substance abuse, they cited their involvement with juvenile justice system more frequently. Interestingly, in all these posts, youth described how their addiction to alcohol and illegal drug consumption led them to commit different petty crimes, e.g., shoplifting, jaywalking, and even DUI offenses, e.g, drunk driving. For instance, a 15-year-old female youth shared a post where she mentioned her drug abuse disorder and several incidents of law violations:
\begin{quote}
\textit{
``\textbf{last night I smoked a shit ton of weed \& the driver (my bestfriends boyfriend) was mad drunk \& trippin off LSD while driving}.. We got pulled over and he got arrested.
I was not supposed to be talking to them because we got arrested for shoplifting and the court said no contact for 6 months and it's only been 2. The probation officer called my mom and I might be on probation for a while. I'm also getting drug tested someday next week so I'm screwed.
''} - 15-year-old female, Juvenile Justice
\end{quote}


\subsubsection{Loss of parents}
Youth also talked about \textbf{losing their parents} in some of the posts (4\%, \textit{n} = 71). As illustrated in \autoref{fig:rq1_chisquared}, the relative proportion of posts related to youth's parents being incarcerated (1\%, \textit{n} = 16) reached positive statistical significance, suggesting that when teens posted about loss of their parents, they also talked about their parents being in prison more frequently compared to other CHINS.  

One of the key trends we observed in these posts was that youth experienced the death of one of their parents, while their other parent faced legal consequences related to drug dealing, fraudulent actions, or abusive behavior, often of a sexual nature towards their children, resulting in the loss of both parents. Interestingly, while most posts were about youth struggling with losing their biological parents, we found a few posts that portrayed the beneficial side of the child welfare system. For instance, a youth shared about her traumatic childhood, but also shared the happiness and closeness they found in her foster family. 

\begin{quote}
\textit{
``
\textbf{My biological mother died from an overdose and my father was put into prison for attempted sexual and physical assault of me.} 
I'm now in a place I feel no one will leave me, living in a house with my gorgeous foster sister and my adopted mum. They're my absolute world I'm so grateful for how things have worked out."
} - 17-year-old female, Parental Imprisonment

\end{quote}
 

\subsubsection{Disability}
In some of the posts (1\%,\textit{n}=23) where youth explained their association with child welfare, juvenile justice, and parental imprisonment,  also included their struggles with their \textbf{mental health issues}. Interestingly, for this dimension of CHINS, we found teens only sharing about their mental and behavioral health problems, e.g., autism, attention deficit disorder (ADD), attention-deficit/hyperactivity disorder (ADHD), bipolar disorder, depression, not any physical disabilities. Also, in majority of these posts, youth cited their involvement with child welfare (1\%, \textit{n} = 19). One of the key trends we observed was that all these youth were physically abused by their biological parents, which often deteriorated their pre-existing mental health issues. What is worse is that their biological families also abandoned them, leading them to be involved in child welfare. For instance, one youth mentioned getting abandoned by their parents because of her neurodiversity:
\begin{quote}
\textit{"\textbf{My real family disowned me for having ADHD.} I like my foster family but it just hurts your family disowning you."} - 16-year-old female, Child Welfare
\end{quote}

Lastly, there were some posts where our youth self-identified themselves as CHINS, but did not mention any specific adverse experiences. Youth who were involved with child welfare had the highest relative proportion of these posts, reaching also a positive statistical significance as shown in \autoref{fig:rq1_chisquared}. To this end, youth implicated in juvenile justice was also found to be statistically significantly associated with these posts while youth whose parents in prison was found to be significantly negatively associated with these disclosures.

In summary, our analysis revealed several insights. We found that youth whose parents were in prisons more likely posted about neglect, emotional abuse, parents' substance abuse, domestic violence, sexual abuse, and loss of parents. We also observed a significant positive correlation between posts describing youth's own substance abuse and their involvement with the juvenile justice system. Lastly, youth who were involved in the child welfare system and posts that discussed self harm showed a significant positive correlation.


\subsection{Topics At-Risk Youth Post About When Seeking Support Online (RQ2)}
\begin{figure}[h]
    \centering
    \includegraphics[scale=0.26]{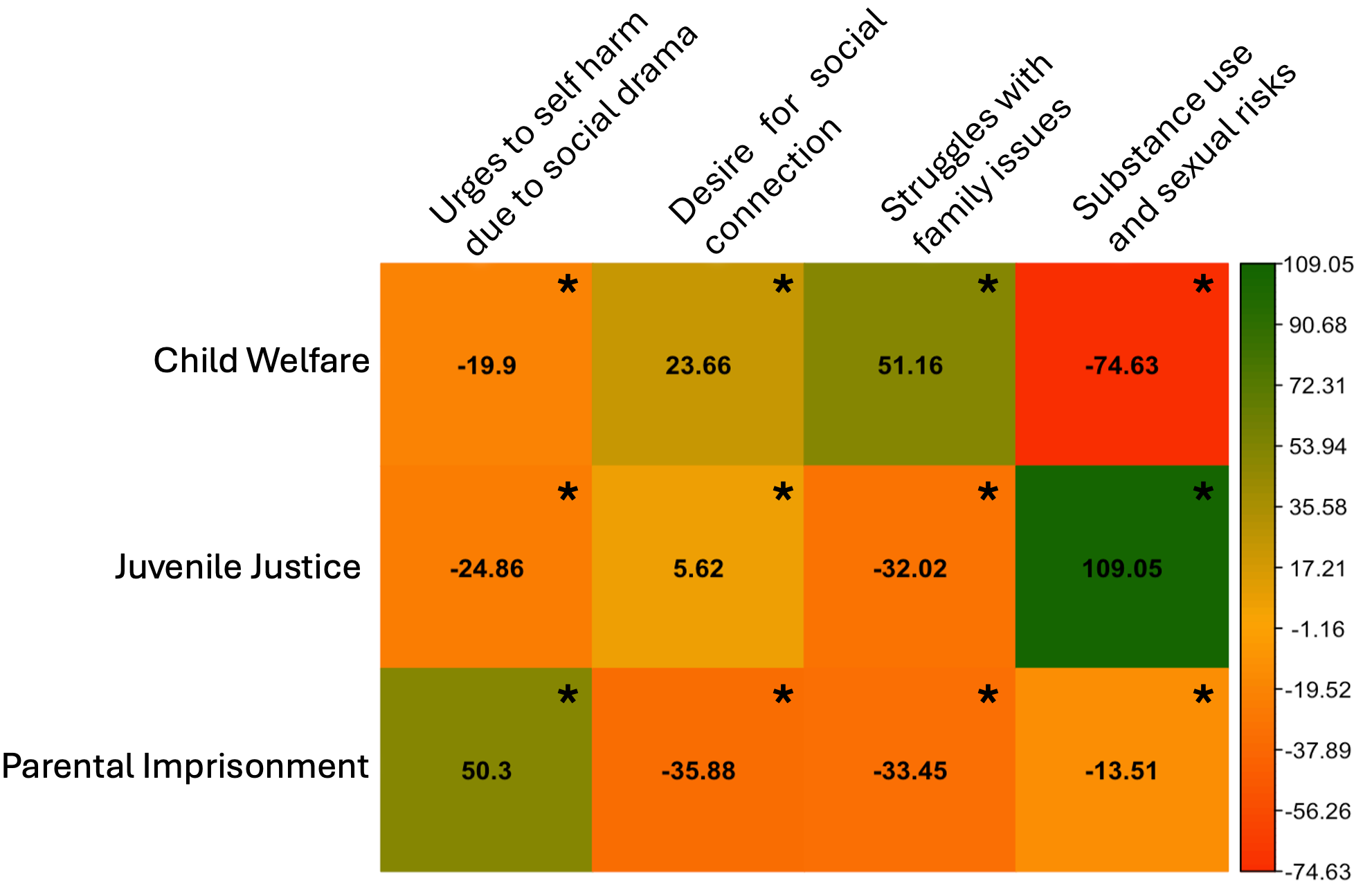}
    \caption{Results (standardized residuals) of the between-group analysis of the top four topics discussed based on the teens' CHINS category ($N=239,929$). (*) indicates significant association. Note that green denotes a positive association, while red denotes a negative one.}
    \label{fig:TP_chisquared}
\end{figure}

\edit{While the above section presented how at-risk youth discussed different adverse childhood experiences (ACES) when they also identified themselves as CHINS, this section sheds light on when CHINS sought support through these posts, what were the most prevalent topics they mentioned.} The four key topics we found were: 
"Urges to Self-Harm due to Social Drama," "Desire for Social Connection," "Struggles with Family Issues," and "Substance Use and Sexual Risks."  The proportion of the four topics discussed by youth in child welfare, juvenile justice, and whose parents are in prison were different (\autoref{table_gamma}). 
Based on a $\chi^2$ test, the proportion of topics discussed differed significantly across the three intervention groups, \(\chi^2\) (6, \textit{N} = 239,929) = 16378.23, \textit{p} <0.001 (\autoref{fig:TP_chisquared}).

The statistically significant trends that we observed include: \edit{
1) youth whose parents are incarcerated were more likely to post about seeking help for their self-harm urges, 
2) youth in the child welfare and juvenile justice system were more likely to share their desire for social connection in their help-seeking posts, 
3) youth in the child welfare system are more likely to post about seeking support regarding their struggles with family, and 
4) youth in the juvenile justice system were more likely to seek support regarding their substance use and sexual risk experiences. 
}
Below, we unpack the four representative topics and how the four topics were discussed among posts shared by teens from three CHINS groups.

\subsubsection{Urges to self-harm due to social drama}

In the majority of the posts (48.5\%, \textit{n} = 116,481), \edit{teens sought support regarding their urges to self-harm due to the various challenges they faced with their social relationships.} Overall, struggles were often linked to their feelings of isolation from social circles due to their familial situations (e.g., being adopted or parents' imprisonment). 
As illustrated in Figure~\ref{fig:TP_chisquared}, \edit{youth whose parents are in prison posted to seek help regarding their self-harming thoughts} proportionately more often than the youth in child welfare or juvenile justice systems. In many posts shared by youth with parents in prison, we observed that they sought peer support explaining their feelings of self-harm because they were bullied at school for their adverse family situation:

\begin{quote}
    \textit{"well iv been upset and afraid for my dad even though i dont really know him but hes addicted to meth and is in jail... but \textbf{i have break downs in the middle of school sometimes cause of all this drama in my life} but my school counciler helps me out a lot. \textbf{i don't know what else to do, y'all should help me.}"} - 14-year-old female, Parental Imprisonment
\end{quote}

In addition, many \edit{posts from} teens whose parents are in prison \edit{revealed a common struggle where} they want to stay clean from self-cutting, yet, they keep having the urge to self-harm because their relationships with friends and/or romantic partners went wrong. 
A similar trend was observed among posts shared by the youth in the child welfare system where \edit{they sought guidance regarding their urges to self-harm because of all the frustrations they felt from their relationships with friends or romantic partners.} 

\begin{quote}
\edit{\textit{"\textbf{So basically yesterday me and my boyfriend split up} and normally I would just cry for a bit and then try get over it all but no this on meant loads to me no matter how much he lied to me about him self I still loved him :/. I couldn't stop crying and a cut all my arm up with a razor :( and now my social worker know and my foster dad!!  \textbf{Hate living this life what should i do?}"}} -15-year-old, female, Child Welfare.
\end{quote}

\edit{In some posts, youth in child welfare even mentioned that they brought razors to cut themselves at school because they felt strong urges to self-harm even during school hours and sought emotional support from their peers. We noticed that, in some of these posts, they were hesitant because they were afraid of being noticed by teachers or friends and stigmatized by them and therefore they sought suggestions to deal with those potential situations.} 
Oftentimes, they were even more scared of being caught by their foster parents about their self-harming behaviors than anyone else because they felt that they would be neglected or punished by their foster parents, rather than supported by them to cope with the current situation. Struggles and frustrations within the family were the most salient issues among the posts shared by youth implicated in the child welfare system.
Meanwhile, in many posts, youth in the juvenile justice system expressed their challenges with their schoolwork, teachers, and friends to the level they want to self-harm. \edit{Many youth expressed self-hatred, calling themselves ``failures'' or ``useless'' or even saying they ``deserved to die'' due to the trouble they had gotten themselves into because of their precarious life situations. The problems just seemed to keep piling up, so that it made it difficult for them to recover. In these cases, they often sought support and advice.} 

\begin{quote}
\edit{\textit{"Got locked up last week. Im 2 weeks behind in school work cause I didn't go for like 2 and 1/2 weeks. \textbf{I'm in a lot of trouble now.} I just can't go, I'm depressed and almost have a panic attack every time I try to go. \textbf{Not going just making me more useless, depressed and anxious. What do I do?"}}} -15-year-old female, Juvenile Justice   
\end{quote}

\subsubsection{Desire for social connection}
\edit{The desire for social connection was the second most representative topic among posts CHINS youth shared online to exchange support and guidance with their peers} (46.9\%, \textit{n} = 112,522).    
Figure~\ref{fig:TP_chisquared} illustrates that \edit{posts from} youth who are in child welfare and juvenile justice are proportionately more likely to \edit{discuss their} wish for social connection, while \edit{posts from} youth whose parents are in prison are less likely to discuss the same topic. Regardless of their familial and/or legal situations, in the majority of the posts, the youth mentioned that they could not sleep because their life was falling apart, and they needed just \textit{anyone} to talk to online, mostly via "Kik." 
An interesting trend we observed was the proactivity of certain youth in extending support to their peers; in many posts, youth in child welfare were not only seeking connections but also offering to listen. 
For instance, a 14-year-old female who had been adopted expressed her willingness to lend a helping hand to those facing challenges similar to her own:

\begin{quote}
   \textit{ ``I'm here to say that I've been going threw what most of you guys are I'm adopted. I'm depressed. I'm suicidal. \textbf{I used to self harm like crazy and I'm here to talk. I've have so my experience I can help I'm just a text away! <3 please do it."}} - 14-year-old female, Child Welfare 
\end{quote}


In many posts, we could see that compared to youth in other CHINS categories, youth in child welfare have been through family struggles for a longer period of time. Being in these long-term and challenging situations made them know how to cope with the challenges, at the same time, they strongly felt the need to connect with online strangers and form romantic relationships with partners partners that could help them cope. \edit{Often times, particularly young women, recounted stories about severe neglect from their family coupled with statements about their current romantic partners, who provide support for them. In the example below, the teenaged girl relied on her promises to romantic partners to help her refrain from self-harm. However, she felt like she let them down when she did not keep her promise and reached out to peers on the platform to get advice on what to do in the aftermath.} 

\begin{quote}
\edit{\textit{“
 At only 4 months old, my Biological father left. Now, I am almost 16 and have not heard a word from him. It never used to bother me, but now that I am older it bugs me [...]. Since my mother was single and always working, I was always with my Grandma and Papa, my great grandfather. \textbf{I just promised both my boyfriends that i wouldnt cut myself, but i did. What the hell do i do?”}}} -16-year-old female, Child Welfare
\end{quote}

A similar trend was observed among the posts shared by teens in the juvenile justice system; teens were mostly about seeking someone to just talk to, mostly on Kik. Unlike posts shared by teens involved in the child welfare system, teens in juvenile systems rarely posted about the offers to listen and help others. Rather, they shared that they are having difficulties with sleeping because of feelings of loneliness and depression.  
Among the posts where teens discuss their needs for social connection, teens were looking for someone to talk to to distract themselves from the bad things happening in their lives. In many posts, teens were looking to make trusted friends online when they felt alone, betrayed, and bullied in their offline lives. These posts re-confirmed the trend discussed above where teens whose parents are in prison are likely to post about frustrations because of social relationships, which triggered urges to self-harm.  

\begin{quote}
    \textit{"\textbf{Someone talk}, I'm dogging school because I'm sick of being bullied and I had to cut. \textbf{My friends never noticed me slip away anyway so they don't care.}"} -14-year-old female, Parental Imprisonment

\end{quote}

\subsubsection{Struggles with family issues}
\edit{The third most representative topic CHINS shared online to look for peer support was their struggles with family issues (2.7\%, \textit{n} = 6,427).} 
\edit{Posts from} youth in the child welfare system discussed their struggles with family issues proportionately more often than the \edit{posts from} youth implicated in juvenile justice systems or whose parents are in prison. Many teens in child welfare posted about the issues with their foster parents and/or siblings in the adopted family (birth children of the adopted parents).
\edit{For instance, some youth sought empathy and compassion from their online peers by posting how they felt that their adopted parents did not show the love or care to them as much they did to their biological children, and did not address the situation fairly when there were conflicts between them and their siblings.}  

\begin{quote}
\edit{ \textit{"I am most likely leaving my home tomorrow. This year might be the last that I ever talk to my adopted parents. \textbf{I am done with how my dad is treating me}. I don't care if I have to live on the streets until I graduate. I am done, \textbf{will someone help me with a place to stay.}"}} -15-year-old male, Child Welfare
\end{quote}

In many of the posts, their struggles with family (both birth family and adopted family) were serious enough to make them hate themselves or self-harm. 
\edit{Furthermore, conflicts with their families drove these youth to seek help and emotional support on online platforms, which occasionally led to encounters with strangers who attempted to exploit their vulnerability. } 
The posts shared by youth in juvenile justice were mostly about the feeling of being rejected by their family members. That included a social comparison between how differently their parents treat their siblings and them. Similar to posts that mentioned urges to self-harm, some teens called themselves the failure of the family when they shared their frustrations and struggles with their family favoring their siblings. 
\edit{Meanwhile, posts shared by teens whose parents are imprisoned were mainly to seek moral support about their struggles with parents and/or family as a whole due to parents' risky behaviors.} In those posts, teens share their efforts and struggles to keep their families together after their parents' imprisonment,
and wanted their peers' suggestions to deal with the situation. 

\begin{quote}
\textit{"I give up so much rn and I can't take it anymore...\textbf{just please I need help rn and I can't think..my mom is about to go to prison}, she was supposed to be here and be my best friend the one till the end and she's gone..."} -14-year-old male, Parental Imprisonment
\end{quote}

\edit{Some teens even share that they wish to be adopted by their extended family rather than continue living with their birth parents who keep being involved in risky and/or illegal behaviors and cannot afford to support the family. As such, unfair treatments received from their birth parents, and craving for a stable life with their family were clearly depicted in the posts shared by youth whose parents are imprisoned.}



\subsubsection{Substance use and sexual risks}
Finally, \edit{the fourth representative topic at-risk youth post about when seeking help online was substance use and sexual risks posed by themselves or their family} (1.9\%, \textit{n} = 4,499).  
Figure~\ref{fig:TP_chisquared} indicates that \edit{posts from teens involved in the juvenile justice system were significantly more likely to seek help about their substance use and sexual risk experiences}, while \edit{posts from} teens in child welfare and whose parents are imprisoned were less likely to do so. 
The majority of the posts shared by youth in the juvenile justice systems were about their experience of substance abuse (e.g., drug abuse, smoking, and drinking) as well as sexually risky behaviors (e.g., having sex in exchange for substance). For instance, \edit{a 16-year-old female youth sought help by posting about his illegal and risky behaviors (i.e. drugs) they were involved in} and how they developed even more serious risks such as sex trafficking. 

\begin{quote}
   \textit{"I recently just got released from juvie. \textbf{Got high last night after partying and having sex with a guy.} I'm coming down off meth and feel like sh*t.. \textbf{any positive words or anything that can help me?} I do not want to use drugs ANYMORE.."} -16 year-old-female, Juvenile Justice 
\end{quote}

\edit{On the other hand, posts from teens with incarcerated parents often showed less inclination toward discussing their own involvement in substance use or sexual risks. Instead, these posts more frequently addressed their parents' risky behaviors and highlighted the potential dangers such behaviors posed to everyone involved, for which they needed advice from others.} For instance, many of the teens expressed their frustration with the situations they can face (e.g., homeless) when their own parents are involved in risky behaviors such as drug overdose.  

\begin{quote}
\textit{"Both of my parents do illegal substances and freak out about me mentioning it do too the fact that they have been arrested before. They throw all there stress/blame on me and I’m tired of it and no longer feel bad for them. \textbf{I need help on deciding weather or not to call the police and if you guys know if there’s a way to call the police anonymously.}"} -17 year-old-female, Parental Imprisonment
\end{quote}

A similar trend was observed \edit{in posts from} teens in the child welfare system, in a slightly different context. \edit{Posts from youth in foster families often sought guidance about their experiences of being sexually abused by the members of the adopted family. For instance, one teen shared her experience about her uncle who adopted her having sexually abused her, and sought online help as she felt she had no one to ask for help around her. }
\begin{quote}
\edit{ \textit{"i am currently 14 years old. when i was 9 i found out that i was adopted by my aunt and uncle.  \textbf{my uncle sexualy abuses me.}  to the point where i have missed so much school because of brusises and cuts on my face i self harm cut burn smoke starve hit scratch all day long dont call me an attention whore like everyone else does \textbf{please i just need help.}"}} -14 year-old-female, Child Welfare
\end{quote}
In summary, when seeking support online, CHINS were not sharing about the everyday problems of typical youth. What they experienced were serious problems that were unimaginable to most people including being neglected by or sexually assaulted by family members. Such adverse life events were observed to traumatize youth at risk to the extent they became strong triggers for their urges to self-harm. 

\subsection{Role of Technology in Exacerbating Adverse Experiences among CHINS (RQ3)}
In this section, we unpack the impact of \textbf{\textit{technology}} in the adverse experiences among CHINS. In the previous section, we revealed that a substantial number of posts made by CHINS were aimed at forging connections with others online. This often led to interactions with strangers online, sharing intimate details and experiences of their lives. Furthermore, we observed a concerning trend in the prevalence of cyberabuse in various forms during online social interaction (e.g., sexting, explicit photo/video sharing, cyber-fraud, cyberbullying, etc.); cyberabuse by family members was not an exception. Another concerning trend was the youth's active involvement in various illegal activities such as substance use and cyber crimes were facilitated by online interaction. Lastly, in a significant number of posts, youth addressed struggles related to self-harm, potentially triggered by exposure to posts recounting similar experiences. Below, we describe the above trends in detail. 

\subsubsection{The sharing of self-harm thoughts and content triggered others:} 
As we examined youth posts, we uncovered a \edit{notable} trend regarding the influence of negative online experiences on the emergence of self-harm thoughts among CHINS. In a majority of the posts, youth highlighted a link between the adverse online encounters and their past traumatic experiences or personal identities (such as neglect, abuse, disability, etc.), which serve as stressors influencing their need to engage in self-harm as a coping mechanism. We discovered that oftentimes, these experiences \edit{were triggered by negative} online interactions with individuals from their social circles, including friends and family. For instance, a 15-year-old girl in the child welfare system shared her experience of getting triggering online messages from her friend which resulted in her urge to engage in self-harm.

\begin{quote}
    \textit{``This girl has been horrible to me for a few weeks now. She calls me a sket, scruff, ugly, fat, retard, thick and \textbf{she texts me horrible stuff and sends me horrible [social media] messages}... \textbf{I self-harm and I even tried killing myself} [...] It's really bad but I dunno what to do cos no one listens to me.''} -15-year-old female, Child Welfare
\end{quote}

Furthermore, our investigation unveiled another striking discovery, as a considerable number of posts detailed instances of retraumatization experienced by CHINS when exposed to content that recounted similar experiences or mirrored their current life circumstances. 

\begin{quote}
    \textit{``I've relapsed. I don't understand. I was doing so well. \textbf{Then I actually triggered myself by looking up videos about suicide and now I'm really depressed.} Why did I do that to myself? Is it possible somewhere like in my subconscious I don't want to get better?''} -17-year-old female, Juvenile Justice
\end{quote}


In summary, our research highlights how the internet can intensify adverse experiences among CHINS. In many cases, their desperate need for social support online made them even more vulnerable to online predators; abusive families never stop harassing them in online spaces; and self-harming thoughts and online content were contagious enough to trigger even more urges to self-harm. In the next section, we discuss the implications of our findings.   

\subsubsection{Seeking social connection led to meeting strangers online:}

In our examination of topics at-risk youth post when seeking support online \textbf{(RQ2)}, we found that many youths were primarily seeking connection with others. However, while youth were seeking social connection, they were susceptible to several forms of risk posed by strangers they met online. One of the reasons for the stranger-danger scenarios they faced was their tendency to share their personal information when seeking help online. In many posts, youth indicated that they are CHINS and willingly disclosed personal information such as their location and mobile phone numbers for the purpose of social connection, underscoring their inclination towards social interaction over privacy concerns.

Sometimes, CHINS were requested to share personal information with strangers online, because in many cases, they indicated that they do not have family or friends who support them in their support-seeking posts. We noticed that youth often struggled with finding the balance between forging connections and safeguarding their privacy online. For instance, a 16-year-old female expressed hesitation to share her phone number with a male stranger while she felt a desire for a social connection with him.  

\begin{quote}
    \textit{
    ``\textbf{This guy is trying to tell me something I'm interested in but I just met him today (online) and he's asking for my number but I just remembered giving your numbers strangers is dangerous.} If he did have bad intentions, what would he do with my number?''} -16-year-old female, Child Welfare
\end{quote}

Additionally, we noticed that as young people tried to make real connections or start romantic relationships online, they often faced unwanted requests for nudes or received inappropriate content from strangers they met online. This trend was quite salient in the posts we reviewed, and it made many youths express their frustration and anger about the situation when seeking help online. For instance, a 16-year-old girl shared her frustration with her inability to engage in normal conversations with guys:

\begin{quote}
    \textit{``\textbf{ughhhh i just want to flirt with some guy without them sending me a dick pic or asking for nudes.} just flirt WITH WORDS.''} -16-year-old female, Child Welfare
\end{quote}

When youth chose to open up and share their emotions through their posts, they occasionally found themselves subjected to various forms of cyberbullying by anonymous individuals online. While youth chose to post as an attempt to seek support, unfortunately, it drew the attention of predators who aimed to exploit the emotional vulnerabilities of these youth. For example, a 17-year-old female reported that guys were wrongly approaching her when she disclosed something personal:
\begin{quote}
    \textit{``\textbf{I posted about being sad, it was personal and suddenly messages from guys trying to hit me up and telling me they're in the "mood."} Honestly, WTF is wrong with people? I cant be a bit vulnerable without the creeps showing up? My profile even says I'm not looking for any of that shit.'' }-17-year-old female, Child Welfare 
\end{quote}

\subsubsection{The internet facilitated cyberabuse by family members and others:}

Our analysis revealed that a significant amount of the posts mentioned their encounters with online harassment, sexting, and cyberbullying. Within this category, youth conveyed various forms of cyberabuse they had experienced. One of the main trends we observed was that many youths reported receiving or getting asked to send intimate content, such as explicit text messages or photos. What was unexpected is that a considerable portion of such inappropriate requests came from individuals within their close social circles, including family members and friends. For instance, a 14-year-old girl shared that she received requests to send her nudes from her close circles including her cousin:

\begin{quote}
    \textit{``I deleted kik...\textbf{Because literally everyone I talked to asked for nudes. Even close friends. Even my cousin.} Now I have less people to talk to. But there's only a few people I enjoy talking to. None of those few people stay up as late as I do. They sleep at night. I feel like I'm being ignored. [...]
    Please help me. Give me some ideas. Please.''} -14-year-old female, Child Welfare
\end{quote}

In different situations involving pre-existing relationships where intimate images had been already exchanged, the dialogue sometimes took a disturbing turn as some individuals resorted to blackmail and threats of publicly posting these pictures when the youth chose to temporarily stop or even end the relationship. For instance, a 16-year-old female mentioned how she was blackmailed after sending nudes to a guy she used to know:

\begin{quote}
    \textit{``So, \textbf{i was sexting this guy I know and then i stopped because i wasn't feeling it anymore and now he's just told me that he's uploaded my nudes online}.. and I'm freaking out.''} -16-year-old female, Parental Imprisonment
\end{quote}

Apart from that, we also observed a notable number of posts addressing experiences related to cyberbullying. What was interesting is that cyberbullying did not solely come from anonymous individuals online, but also from people within the youth's own families. In these specific situations, the act of posting content online to seek support by youth exposes their vulnerabilities, resulting in them becoming targets of cyberbullying on the platform by members of their own families. For instance, a 13-year-old girl in child welfare talked about how her own sister was bullying her after her sister saw her username of the mobile app on her phone without her consent:

\begin{quote}
  \textit{``\textbf{My sister went on my phone and seen this app.} Now she knows my username. \textbf{Now she likes to go through my posts and call me, "Sad, Emo madi" and mock me saying, "awww" sarcastically.}''} -13-year-old female, Child Welfare
\end{quote}

\subsubsection{The Internet facilitated illegal and risky behavior:}
\edit{In our analysis of teens' posts, we found that some online social interactions had the potential to expose some} CHINS to illicit and illegal activities. A substantial number of posts narrate previous experiences or the active involvement of CHINS in risky pursuits, such as seeking advice on the usage of illegal substances, or encouraging and enticing fellow teenagers to partake in perilous endeavors. Among these posts, the most prevalent risk was active seeking of social connections and assistance from other users to engage in illegal activities (such as underage drinking, substance abuse, etc.). For instance, a 14-year-old female was soliciting suggestions on illegal drugs:

\begin{quote}
    \textit{``I've currently taken a liking to drugs, but I'm out of molly and weed.. :/ \textbf{does anyone know some over the counter meds I can use?}''} -14-year-old female,  Parental Imprisonment
\end{quote}

There was also a trend regarding the involvement of CHINS in romantic relationships characterized by risky behavior with individuals they have met online. For instance, a 16-year-old female, who identified herself as adopted, recounted her personal experience of meeting someone she had initially encountered online in person, leading to risky sexual behaviors and a potentially challenging situation of a possible pregnancy. 
\begin{quote}
    \textit{``\textbf{I met a guy on here and he drove to see me. Im 16 and hes 27 and the plan was for him to get me pregnant so i could go be with him and leave my adopted parents place.} He came last month and we had unprotected sex a few times but now hes gone [...] now im freaking out and could be pregnant. Idk what to do''} -16-year-old female, Child Welfare
\end{quote}

 \edit{In summary, CHINS who already experienced significant adverse life events used the internet to seek help and support, but often times their vulnerability exposed them to even more trauma instead of playing a protective role in helping them overcome the adversity they faced.} 

 \section{Discussion}
In this section, we describe the implications of our findings in relation to prior work and offer design guidelines for creating online support platforms tailored to CHINS. \edit{Overall, our findings underscore the importance of designing online platforms that address CHINS' unique struggles related to their ACE's, promote safety and resilience, and provide tailored support to mitigate risks and foster well-being.}

\subsection{The Adverse Experiences of CHINS Provide Insight into Their Unique Struggles (RQ1)}

In our analysis, identifying CHINS was relatively straightforward and provided valuable insights into their offline and online challenges. While this identification process was beneficial for our research, a concern arises that the same approaches we used could be exploited by individuals with malicious intentions \cite{PDFProte87:online}. This emphasizes the need to offer CHINS education and awareness on privacy and security practices, empowering them to be less vulnerable to online predators. Researchers, developers, and educators could work together to directly engage with at-risk youth to establish proactive safeguarding and preventive measures, thereby raising awareness among potential young victims. \edit{Previous research on networked privacy and related topics advocates for user-centered privacy educational approaches, not just restricting disclosures (abstinence-focused interventions) \cite{Smith2023PIE, Tazi2023CyberOnlineEd}. A user-centered approach to privacy education would address risk and awareness at multiple levels of a child's socio-ecological systems \cite{badillo2024towards} by including various teaching methods (e.g., lecture-based learning, technology-based learning, group and individual learning) as well as the participation of various stakeholders (e.g., children, parents, foster parents, caseworkers, teachers)~\cite{park2024resilience, park2023autonomy, wisniewski2025moving, caddle2024stakeholders}.} 

We found that these youth were more open about their ACEs, which differed significantly from typical issues within the general youth population. These challenges often originated from traumatic experiences faced by CHINS in their present or past lives, which may not be suitable to share with youth who have not experienced similar trauma. Tajfel's social identity theory \cite{tajfel1979integrative} suggests that group identity, such as identifying as a CHINS, strongly influences an individual's self-esteem. Group identity is considered an extension of self-concept and can impact attitudes, behaviors, information appraisal, and help-seeking behaviors \cite{KLIK201935}. Thus, creating smaller, safer, and moderated communities based on youth identities (e.g., CHINS) rather than experiences (e.g., ACES) may be more effective in garnering support \cite{ren2007applying}. \edit{Although all of these youths required services, the three CHINS categories faced distinct adverse experiences that contributed to varying challenges in their lives.  As support groups are often aligned with youth identities, it's crucial that social services are also tailored to address their specific adverse childhood experiences (ACEs) to prevent re-traumatization when they self-identify and share their experiences online. Hence, we advocate for the development of tailored trauma-informed approaches to address their specific needs. Trauma-informed approaches have been explored in the design of various digital systems and social platforms in HCI \cite{chen2022trauma, razi2024toward, randazzo2023trauma, scott2023trauma}}. This approach adheres to six core principles: 1) Safety, 2) Trustworthiness and Transparency, 3) Peer Support, 4) Collaboration and Mutuality, 5) Empowerment, Voice and Choice, 6) Cultural, Historical, and Gender Issues. We can apply a trauma-informed approach in the context of CHINS by designing technologies and supports that incorporate these principles. For example, research investigations should center the lived experiences of CHINS through their own stories and data (such as this paper) to amplify the voices of these youth and bring to the forefront their needs. Furthermore, researchers and designers should prioritize partnership, mutual power, and shared decision-making, using methods such as participatory design \cite{xiao2022sensemaking, chatlani2023teen, ali2024ideas, agha2024tricky, jean2023teens} to include CHINS in the design process of technologies built to protect them.

\edit{\subsection{CHINS Seek Online Support; Meanwhile, the Internet also Facilitated Additional Risks and Adverse Life Experiences (RQ2 \& RQ3)}}

\edit{Our research demonstrates the internet's dual nature, showing how CHINS are actively seeking support via the internet, while the internet also serves as an amplifier of their adverse life experiences. Our findings showed that CHINS sought social connection, often stemming from challenges within their families. At the same time and while not the primary cause of trauma, the internet facilitated and intensified the impact of the CHINS' adverse experiences, with instances of CHINS being retraumatized by online content, leading to self-harm tendencies. Prior research by Badillo-Urquiola et al.~\cite{badillo2019risk} highlighted that teens in the child welfare system may feel rejection, leading to a desire for acceptance and love. Adverse experiences shared by CHINS in our study indicated a lack of strong ecological support networks, prompting them to seek advice and support online, resulting in potential online risks. Furthermore, past studies noted the contagion of self-harming behaviors among youth through online sharing \cite{khasawneh2020examining, abraham2022applying}. Our findings indicate heightened susceptibility in at-risk youth, given their traumatic experiences. Therefore, peer support mental health platforms, especially if not moderated by professionals, may not provide the healthiest support outlet for CHINS. We advocate for providing proactive measures for vulnerable youth to shield themselves from exposure to online self-harming content, preventing further trauma. Yet, these measures must carefully balance between protecting the youth and allowing them to participate in beneficial peer support \cite{ali2024m}. By definition, CHINS have higher support needs that require more social ecological support at all levels of their social networks, such as individual (self), family (parents, foster parents), community (Guardian Ad Litems, Church, Nonprofits), and societal (policies, government), to meet their higher needs. Such a collaborative approach \cite{akter2023COoPS, akter2023evaluation}, involving resources and education targeted to individual needs \cite{wright2020child, bronfenbrenner1979ecology}, can foster ongoing relationships between youth, parents \cite{akter2022parentteen, akter2023extended, akter2024towards}, caregivers \cite{kropczynski2021caregiving}, caseworkers, and mentors \cite{badillo2019risk}. Open dialogues can improve understanding of youth challenges, nurturing trusting family and community relationships. Additionally, strength-based training, such as co-parenting management for divorced parents, can also provide families with essential support, reducing the vulnerability of CHINS when seeking help from strangers online.}

\edit{CHINS in our dataset shared deeply traumatic and adverse life experiences that often led to disclosing personal information that further exposed their vulnerability. Their motivation for seeking connection and belonging, while understandable, exposed them to online risks like online grooming and cyberabuse. Although online friendships are common among youth \cite{Pew2018}, nearly 90\% of sexual advances toward youth occur online \cite{Childcrimeprevention_2023}. Furthermore, about 1 in 6 youth in foster care falls victim to sex trafficking during runaway episodes \cite{Runaway2020}. Our findings, alongside established literature on human trafficking, highlight these youth's heightened vulnerability to sexual predation. Awareness campaigns and tools to combat child sexual abuse are imperative \cite{Thorn2023}. In addition, rather than restricting or overly controlling CHINS' online access in fear of risks \cite{risk_narrative}, asset-based approaches \cite{badillo2020assets} should be implemented to teach them positive digital skills. For instance, mental health platforms can use nudges and safety features to support better decision-making and reduce exposure to online risks \cite{agha2023strike}. Overall, interventions for CHINS should be tailored to their heightened risk experiences and needs, rather than viewing it in one-size-fits-all manner, emphasizing targeted risk prevention approaches \cite{Alsoubai2024Profiling}.  Our results provide a basis for developing support systems tailored to the unique needs of CHINS, as discussed next. }\\

\subsection{Implications for Design}
Designing online support platforms for CHINS requires a nuanced approach that considers their adverse life experiences and current struggles tied to their identity. \edit{Drawing from our findings,} we outline below practical design implications for addressing the unique needs of CHINS in peer support platforms:
    \subsubsection{Provide safe places for seeking support:} We encountered posts where CHINS shared instances of online risks they encountered, such as exposure to harmful content, sexual solicitations, and bullying (even by their own family members), when seeking support on peer-support platforms. \edit{CHINS in our study also mentioned that seeing posts about the adverse experiences of others often triggered hurtful memories or urges to self-harm.} Hence, designers need to focus on implementing safeguards and ways of evaluating the quality of support they receive from online peer-support platforms~\cite{hartikainen2021safe}. For instance, voting systems combined with professional expert opinion can be used to ensure the safety of CHINS when seeking support. To discourage anti-social behaviors online such as cyberbullying, and sexual solicitations, user-controlled priority flagging options can be incorporated so that CHINS can moderate their online interaction and make urgent reports for immediate attention in critical situations. While a limitation to this approach could be possible cases of misuse by users~\cite{hartikainen2021safe}, timely feedback mechanisms updating users on their reports could be useful to ensuring transparency and responsible use of the flagging option among users. \edit{Providing opportunities to flag unwanted topics or content could also help CHINS avoid triggering posts.}

    \subsubsection{Real-Time Self-Harm Prevention:} \edit{CHINS in our study often shared posts describing their reliance on self-harm as a way of coping with past and/or current traumatic experiences. When seeking support online, they often shared instances of stigmatization and microaggression from people within their social circle making it difficult to seek help for self-harm and in some cases, causing a relapse to self-harm. To address these findings,} designers should prioritize real-time technological interventions that can foster a safe and supportive environment for CHINS to seek assistance. To provide well-tailored support to prevent self-harm, features such as anonymous helplines or chats can be implemented in peer-support platforms to help CHINS who feel discomfort seeking help from adults and others within their social circle. Given the limited parental involvement or experiences of neglect among CHINS, incorporating advanced self-harm monitoring algorithms \cite{stewart2020risk, scherr2020detecting} and content moderation algorithms \cite{saxena2023} with dynamic customization options based on CHINS' preferences and comfort levels, can potentially create personalized and supportive experiences within peer-support platforms. These algorithms can be useful to identify indicators of self-harm or distress within posts, triggering automatic alerts to moderators or trained peer support counselors. Such proactive approaches can promote timely and supportive responses to mitigate potential risks and foster the well-being of at-risk youth. Additionally, ensuring the availability of 24/7 online therapy sessions and self-esteem-building tools such as affirmation reminders in peer-support platforms can help deliver positive affirmations to boost CHINS' self-esteem and emotional resilience, potentially aiding their healing journey from adverse experiences.

    \subsubsection{Automated Risk Detection \edit{and Nudges}:} \edit{Our findings highlight that CHINS seek support for their adverse experiences on online platforms. However, CHINS may also be sharing too much personally identifiable information online that could make them more susceptible to online risks.
    Designers and developers can consider implementing risk detection algorithms in the form of nudges to guide at-risk youth in privacy protection practices and} better decision-making by providing feedback on possible risky online interactions~\cite{agha2023strike}. Rather than enforcing detection algorithms as a means of policing online activities of youth~\cite{caddle2023duty}, they should be used as tools to promote emotional resilience among CHINS. Emotional resilience are well-tailored strategy employed to help CHINS cope with adverse emotional traumas and navigate challenging situations. For instance, peer-support platforms can promote emotional resilience through alerts or nudges, notifying CHINS about the consequences of sharing personal information or engaging in risky online interactions. Incorporating teachable moments as educational opportunities can empower CHINS by fostering positive coping mechanisms, improving emotional management \cite{sylvia2022}, and reducing the risk of re-traumatization. Frameworks such as behavioral analytics~\cite{alsoubai2022mosafely} could be integrated into peer-support platforms to identify risky behavior patterns and be able to notify Social Service Providers (SSPs) in a timely manner \cite{xavierDuty2022, Agha2021Justintime}. To combat substance abuse exposure, designers can incorporate automated content filters \cite{wisniewski2017parental} as safeguards, reducing access to inappropriate content or platforms that may encourage illicit activities and the distribution of illegal substances.

    \subsubsection{Online Resource Hub:} \edit{Based on our findings, youth with incarcerated parents often discussed domestic violence and physical abuse when seeking support. As such, we recommend that online peer support platforms} should prioritize dedicated pages containing resources for domestic violence \cite{storer2023technology}, featuring emergency contacts, shelter details, and guidance on seeking help. Considering the prevalence of sexual abuse cases shared by CHINS online, platforms should also offer resources, professional help, and confidential reporting mechanisms \cite{razi2020let, namuggala2023social}. For example, online counseling services \cite{campbell2018preliminary} with therapists specializing in trauma, abuse, and ACEs could be incorporated into the resource hub to effectively address CHINS' questions and concerns.
   We observed support seeking posts in which CHINS described their involvement in inappropriate media sharing and sexting. To address this, integrating online gamified safety campaigns and digital citizenship programs into online support platforms and social media on-boarding guidelines is crucial. These initiatives will help educate CHINS on responsible online behavior, ethical technology use, and the significance of creating positive digital footprints. Covering topics such as online etiquette \cite{bayor2018}, responsible social media usage \cite{dennen2019}, and consequences of online actions \cite{franzo2019}, these programs can play a vital role in enhancing CHINS' awareness and skills. Online mentorship programs \cite{konkel2016life} can also be an option on peer-support platforms to steer at-risk youth toward constructive life choices by connecting them with mentors who provide valuable guidance and help prevent them from being misled online.

\subsection{Limitations and Future Research}
There are several limitations of our study that can inform future research. Firstly, given that the online peer-support platform was geared towards youth and young adults with mental health struggles \cite{tanni2024lgbtq}, \edit{which may not represent the diversity of online platforms or the experiences of all at-risk youth.} It is possible that our findings may be biased toward CHINS who are struggling, rather than those who are thriving. Therefore, while the picture painted in this paper is bleak, there is a chance that the adverse experiences of some CHINS may not be as grim as what is reported here. While this is our hope, we argue that research on vulnerable populations who are in need of support is a worthy endeavor to continue pursuing. At the same time, future research should consider studying CHINS who have managed to bring themselves or who have support networks that have helped bring them out of adversity. Such research wouldhelp in identifying protective factors for CHINS more broadly, so that they can overcome the trauma from their childhood, rather than continue to relive it. \edit{Next, while our research provides a cross-sectional snapshot of online behaviors of CHINS and how they seek support for their ACEs, future research should consider a longitudinal view of how these behaviors and risks evolve over time for at-risk youth.} Furthermore, not all CHINS may have access to the internet and peer support platforms, given prior research has shown that adults may restrict access to technology in the home, particularly for foster youth, as a means to protect them \cite{badillo2019risk}. Therefore, future research should also consider the needs of CHINS who do not have access to online support systems. Finally, our work does not provide insight into the types of support and advice CHINS receives via online peer support platforms. That is an area of future work we plan to undertake based on the comments made on the posts analyzed in this paper.

\section{Conclusion}
\edit{This paper presents the lived experiences shared by youth implicated in the child welfare and juvenile justice systems or who have a parent in prison. Our findings demonstrate that these CHINS are desperate for support and reaching out to strangers on the internet to help them. As technology researchers, designers, and developers, we must amplify their voices and create online experiences that are tailored to their needs to promote safety and resilience.}

\begin{acks}
Dr. Wisniewski's work on adolescent online safety is supported in part by the William T. Grant Foundation Award \#187941 and the National Science Foundation Award IIS-2333207. Dr. Badillo-Urquiola's work on online safety for foster youth is partially supported by an unrestricted gift from Google. Any opinions, findings, conclusions, or recommendations expressed in this material are those of the authors and do not necessarily reflect the views of our sponsors.
\end{acks}

\bibliographystyle{ACM-Reference-Format}
\bibliography{References}

\newpage
\appendix
\section{Acronyms and Definitions}

\begin{table}[htbp]
\caption{Acronyms and Definitions}
\label{tab:acronyms}
\centering
\footnotesize
\begin{tabularx}{\linewidth}{p{3cm}|X}
\hline
\rowcolor[HTML]{EFEFEF}
\textbf{Acronyms} & \textbf{Definitions} \\ \hline
CHINS & Children in Need of Services \\ \hline
ACES & Adverse Childhood Experiences \\ \hline
CPS & Child Protective Services \\ \hline
DCF & Department for Children and Families \\ \hline
DCS & Department of Child Services \\ \hline
CYF & Children, Youth and Family \\ \hline
DCFS & Department of Children and Family Services \\ \hline
DSS & Department of Social Services \\ \hline
DHS & Department of Human Services \\ \hline
DHR & Department of Human Resources \\ \hline
\end{tabularx}
\end{table}

\newpage
\begin{table*}[htbp]
\raggedright
\scriptsize
\caption{CHINS Criteria and ACES Mapping. Note: Blank table cells mean that the criteria were mutually exclusive to either CHINS or ACES.}
\label{tab:mapping}
{\renewcommand{\arraystretch}{1.5}
\begin{tabular}{|p{2cm}|p{4cm}|p{6cm}|p{1cm}|}
\hline
\rowcolor[HTML]{EFEFEF}
\textbf{Definitions} & \textbf{Children in Need of Services (CHINS)} & \textbf{Adverse Childhood Experiences (ACES)} & \textbf{Codes} \\ \hline
\rowcolor[HTML]{FFFFFF}
Neglect & Well-being is seriously harmed because their parent, guardian, or custodian is unable, unwilling, or neglectful in providing care. & 
\begin{itemize}[leftmargin=*]
    \item No one in your family loved you or thought you were important or special?
    \item Your family didn’t look out for each other, feel close to each other, or support each other?
    \item You didn’t have enough to eat, had to wear dirty clothes, and had no one to protect you?
    \item Your parents were too drunk or high to take care of you or take you to the doctor if needed.
\end{itemize}
& NEG \\ \hline
\rowcolor[HTML]{EFEFEF}
Sexual Abuse & Victim of sexual abuse, sexual trafficking, or sexual-related offenses. & 
Did an adult or person at least 5 years older than you ever touch or fondle you, or have you touch their body in a sexual way? Or attempt or actually have oral or anal intercourse with you? 
& SEX \\ \hline
\rowcolor[HTML]{FFFFFF}
Physical Abuse & Physical harm or endangered due to injury caused by actions or negligence of their parent, guardian, or custodian. &
\begin{itemize}[leftmargin=*]
    \item Did an adult or person at least 5 years older ever act in a way that made you afraid you might be physically hurt?
    \item Did a parent or adult in the household often push, grab, slap, or throw something at you? Or ever hit you so hard that you had marks or were injured?
\end{itemize}
& PHY \\ \hline
\rowcolor[HTML]{EFEFEF}
Emotional Neglect / Abuse & Emotional neglect or abuse is causing harm to the child's emotional and psychological development. & 
Did a parent or other adult often swear at you, insult you, put you down, or humiliate you? 
& EMO \\ \hline
\rowcolor[HTML]{FFFFFF}
Domestic Violence & The child is exposed to domestic violence within their household, which may adversely affect their well-being. & 
\begin{itemize}[leftmargin=*]
    \item Was your mother or stepmother often or very often pushed, grabbed, slapped, or had something thrown at her?
    \item Sometimes, often, or very often kicked, bitten, hit with a fist, or hit with something hard? Or ever threatened with a gun or knife?
\end{itemize}
& DOM \\ \hline
\rowcolor[HTML]{EFEFEF}
Substance Abuse (Self) & The child is involved in the use or distribution of illegal substances, drug abuse, underage drinking, or similar activities. & & SUBS \\ \hline
\rowcolor[HTML]{FFFFFF}
Substance Abuse (Parent) & Parent or guardian is engaged in substance abuse or related activities, which can negatively impact the child. & 
Did you live with anyone who was a problem drinker or alcoholic, or who used street drugs? 
& SUBP \\ \hline
\rowcolor[HTML]{EFEFEF}
Self-harm/ Imminent Risk (Self) & The child poses a risk to their own well-being through self-harming behavior or actions that may result in imminent harm. & & MENS \\ \hline
\rowcolor[HTML]{FFFFFF}
Self-harm/ Imminent Risk (Family) & A family member's involvement in self-harming behaviors or other imminent risks. & 
Was a household member depressed, mentally ill, or did they attempt suicide? 
& MENS \\ \hline
\rowcolor[HTML]{EFEFEF}
Disability & The child has a disability and lacks access to necessary medical interventions and support. & & DIS \\ \hline
\rowcolor[HTML]{FFFFFF}
Loss of a Parent & The child has experienced the loss of one or both parents. & 
Was a biological parent ever lost to you through divorce, abandonment, or other reason? 
& LOSP \\ \hline
\rowcolor[HTML]{EFEFEF}
Foster Care & The child has been placed in the foster care system or has had involvement with child welfare services. & & CW, FOS, AD \\ \hline
\rowcolor[HTML]{FFFFFF}
Child Protective Service & The child has come into contact with social services due to their own actions or those of their parents or guardians. & & CW, FOS, AD \\ \hline
\rowcolor[HTML]{EFEFEF}
Juvenile Justice & The child has been involved in criminal activities or has had contact with juvenile detention centers. & & JJ \\ \hline
\rowcolor[HTML]{FFFFFF}
Legal System & The child is currently or has previously been involved in a legal process. & & JJ \\ \hline
\end{tabular}
}
\end{table*}

\newpage
\clearpage
\section{RQ1 Codebook}
\begin{table*}[htbp]
 \centering
 \scriptsize
 \caption{RQ1 Codebook. Note: percentages in the codes column are based on total posts (\textit{N} = 1,663).}
   \label{tab:codebook}
\begin{tabular}{ |p{1.8cm}|p{3.3cm}|p{8.4cm}|  }
\hline {\textbf{Codes}} & {\textbf{Risk Category}} & {\textbf{Illustrative Quotations}} \\ 

 \hline
 \rowcolor{lightgray}
 \multicolumn{3}{|c|}{} \\[-8pt] 
 \multicolumn{3}{|c|}{\textbf{Self-Harm (Self and Family)}}  \\ \cline{1-3}
Self-harm (MENS) \newline 22\%, \textit{n} = 370  &
Child Welfare (\textit{n} = 314) \newline
Juvenile Justice (\textit{n} = 21) \newline 
Parental Imprisonment (\textit{n} = 35) & 
\textit{“I’m adopted. I’m a severe self harmer. nothing seems to go right in my life, and I’m only 13. I don’t want to live, I have severe depression and I don’t know what to do. I feel like I’m crazy but I can’t help it."} -14-year-old female, Child Welfare \\ \cline{1-3}
Self-harm in family (MENF) \newline 17\%, \textit{n} = 17  &
Child Welfare (\textit{n} = 13) \newline
Juvenile Justice (\textit{n} = 1) \newline 
Parental Imprisonment (\textit{n} = 3) & 
\textit{"[...] to live with my grandmother I became very depressed and started cutting and starving myself like eating 2 meals a week. So good bye world. I never understood why I was put here in the first place it’s been 15 years of pure hell guess I won’t ever figure it out good bye cruel world."} - 16-year-old female, Child Welfare  \\ \cline{1-3}

 \hline
 \rowcolor{lightgray}
 \multicolumn{3}{|c|}{} \\[-8pt] 
 \multicolumn{3}{|c|}{\textbf{Neglect and Abuse}}  \\ \cline{1-3}
Neglect (NEG)\newline 16\%, \textit{n} = 262  &
Child Welfare (\textit{n} = 210) \newline
Juvenile Justice (\textit{n} = 6) \newline 
Parental Imprisonment (\textit{n} = 46) & 
\textit{"The worst thing about being adopted into a family is you never feel like you’re a part of it, and you know you were never supposed to be a part of it. You know you don’t belong, you know they love their kids more than you, you know they’d rather be without you."} - 16-year-old female, Child Welfare  \\ \cline{1-3}
Physical Abuse (PHY) \newline 6\%, \textit{n} = 101  &
Child Welfare (\textit{n} = 79) \newline
Juvenile Justice (\textit{n} = 7) \newline 
Parental Imprisonment (\textit{n} = 15) & 
\textit{"When I was about 2 years old my mother and father had beat me all the time. My dad scarred my hand with a curling iron. This scar is extremely noticeable, and I have had many questions asked. After their abuse and alcohol addiction, I was ripped from my parents by the police at age 4. Then I was adopted into a new family with better parents. They are about average."} - 13-year-old male, Child Welfare  \\ \cline{1-3}
Emotional Abuse (EMO)\newline 4\%, \textit{n} = 65  &
Child Welfare (\textit{n} = 58) \newline
Juvenile Justice (\textit{n} = 4) \newline 
Parental Imprisonment (\textit{n} = 3) & 
\textit{"Whenever she says I never do enough of something I go to the effort of doing things better but she still yells at me saying I don’t love and respect them... I ask if I can help around the house I ask if I can do anything for them. They always reply "no" " She keeps saying "you know what I want you to do so why am I still yelling at you? you’ve been acting weirdly even since you were 6 years old we never gained anything from you". She says there’s something mentally wrong with me... I don’t know what I’m missing out on... "} - 17-year-old female, Child Welfare  \\ \cline{1-3}
Domestic violence (DOM)\newline 4\%, \textit{n} = 60  &
Child Welfare (\textit{n} = 34) \newline
Juvenile Justice (\textit{n} = 7) \newline 
Parental Imprisonment (\textit{n} = 19) & 
\textit{"I’m in hospital with broken arm leg and ribs.also a small bleed to the brain Got an operation in the morning to fix the bleeding and put metal plates in leg and arm. Dad n bro arrested."} - 14-year-old female, Parental Imprisonment  \\ \cline{1-3}
Sexual abuse (SEX)\newline 8\%, \textit{n} = 127  &
Child Welfare (\textit{n} = 97) \newline
Juvenile Justice (\textit{n} = 3) \newline 
Parental Imprisonment (\textit{n} = 27) & 
\textit{"So, i was sexually assaulted by my father multiple times when i was eight and younger. He applied for parole a month ago but didn’t get it. Im frustrated because ...I want to go out and have fun but I keep having flashbacks of the assault.[...] I can see everything happening in my mind almost as though it were happening now. I just want to go out and have fun but i cant. Advice anyone?"} - 13-year-old transgender, Parental Imprisonment  \\ \cline{1-3}
 \hline
 \rowcolor{lightgray}
 \multicolumn{3}{|c|}{} \\[-8pt] 
 \multicolumn{3}{|c|}{\textbf{Substance Abuse of Self or Family Members}}  \\ \cline{1-3}
Substance abuse-Parent (SUBP) \newline 6\%, \textit{n} = 108  &
Child Welfare (\textit{n} = 71) \newline
Juvenile Justice (\textit{n} = 3) \newline 
Parental Imprisonment (\textit{n} = 34) & 
\textit{"When I was five I was molested by my sisters dad my mom was a druggy and sometimes did it in front of us and mentally verbally and physically abused my little sister and I, then at the age of 7 I was raped and molested and forced to do disgusting things till the age of 8 by two different people then DCF took me away when I was in 4th grade cause my mom was doing drugs and I was raising myself and my little sister then in 5th [...] bye world I never understood why I was put here in the first place it's been 15 years of pure hell guess I won't ever figure it out good bye cruel world " - 16-year-old, female, Child Welfare}  \\ \cline{1-3}
Substance abuse-Self (SUBS) \newline 3\%, \textit{n} = 50  &
Child Welfare (\textit{n} = 36) \newline
Juvenile Justice (\textit{n} = 13) \newline 
Parental Imprisonment (\textit{n} = 1) & 
\textit{"Im not sure what to do, last night was crazy. I smoked a shit ton of weed \& the driver (my bestfriends boyfriend) was mad drunk \& trippin off LSD while driving.. We got pulled over and he got arrested and I was the youngest there, they were all 17 \& I was only 14.. Let alone I’m not supposed to be talking to my bestfriend because we got arrested for shoplifting and the court said no contact for 6 months and it’s only been 2.. The probation officer called my mom and I might be on probation for a while, I’m also getting drug tested someday next week so I’m screwed."} - 15-year-old female, Juvenile Justice  \\ \cline{1-3}

 \hline
 \rowcolor{lightgray}
 \multicolumn{3}{|c|}{} \\[-8pt] 
 \multicolumn{3}{|c|}{\textbf{Loss of Parents}}  \\ \cline{1-3}
Loss of parent (LOSP) \newline 4\%, \textit{n} = 71  &
Child Welfare (\textit{n} = 54) \newline
Juvenile Justice (\textit{n} = 1) \newline 
Parental Imprisonment (\textit{n} = 16) & 
\textit{"I feel this year I’ve found my "family". My biological mother died from an overdose and my biological father was put into prison for attempted sexual and physical assault of me when he reached out to meet me after the years he betrayed me. I’m now in a place I feel no one will leave me, living in a house with my gorgeous foster sister and my adopted mum."} -17-year-old female, Parental Imprisonment \\ \cline{1-3}

 \hline
 \rowcolor{lightgray}
 \multicolumn{3}{|c|}{} \\[-8pt] 
 \multicolumn{3}{|c|}{\textbf{Disability}}  \\ \cline{1-3}
Disability (DIS)\newline 1\%, \textit{n} = 23  &
Child Welfare (\textit{n} = 19) \newline
Juvenile Justice (\textit{n} = 1) \newline 
Parental Imprisonment (\textit{n} = 3) & 
\textit{"My real family disowned me for having ADHD. I like my foster family but it just hurts your family disowning you."} - 16-year-old female, Child Welfare  \\ \cline{1-3}

\end{tabular}
\end{table*}

\newpage
\section{RQ2 Topics and Distributions}
\begin{table*}[htbp]
\caption{Top Four Topics and Example Quotes (RQ2)}
\label{tab:fourtopics}
\footnotesize
\begin{tabularx}{\textwidth}{p{4.5cm}|X}
\hline
\rowcolor[HTML]{EFEFEF}
\textbf{Topics} & \textbf{Example Quotes} \\ \hline
\textbf{Urges to Self-harm Due to Social Drama} & \textit{"I'm having a lot of issues with my schooling life. I'm afraid of my future and want to prevent it from happening. I cut. I hate myself. There's a burning in my stomach and an empty whole in my chest. I want to die."} \newline \textit{"I'm losing my best friend because I don't want to do drugs, and drink, and party anymore. It's her birthday today so I'm about to drink. And soon I'm going to hate myself for it and start cutting again. Back to the routine."} \\ \hline
\textbf{Desire for Social Connection} & \textit{"I have never really done this but dose anyone wanna kik I can't sleep [...] My kik is (username)"} \newline \textit{"Need to talk? I've been through more than people think I have. I'm up for anything! Kik me at (username). There has never been a question I haven't answered. And I ALWAYS reply :)"} \\ \hline
\textbf{Struggles with Family Issues} & \textit{"I hate my life my foster mother calls me fat almost everyday and she doesn't know it hurts I have a very low self-esteem now ever since I got adopted and she's always asking what's wrong with me well I wonder what."} \newline \textit{"Sitting home right now with my siblings because my mom got arrested last night for DUI. She does not listen... I tell her not to drive drunk. And now we don't have enough money to get her out."} \\ \hline
\textbf{Substance Use and Sexual Risks} & \textit{"Okay so I was on drugs for a while didn't have the money to pay for them. I ended up sleeping with guys in exchange for them."} \newline \textit{"Fuck my mom, I can't wait until I move out. I fucking know she's doing drugs again. She's an idiot. She knows she'll go to prison if she messes up again."} \\ \hline
\end{tabularx}
\end{table*}

\begin{table}[htbp]
\footnotesize
\caption{Distribution of the Four Topics in Three CHINS Categories in Percentage (approximate number of posts) (RQ2)}
\label{table_gamma}
\begin{tabular}{@{}llllll@{}}
\toprule
                      & Self-harm     & Social Connection & Family Struggle & Risk Behavior & Total         \\ \midrule
Child Welfare         & 47.2\% (78,504) & 48.5\% (80,666)     & 3.8\% (6,321)     & 0.5\% (832)     & 100\% (166,323) \\
Juvenile Justice      & 42.7\% (15,665) & 48.4\% (17,744)     & 0.02\% (73)       & 9\% (3,300)     & 100\% (36,772)  \\
Parental Imprisonment & 60.9\% (22,322) & 38.5\% (14,112)     & 0.09\% (33)       & 1\% (367)       & 100\% (36,834)  \\\hline
Total                 & 48.5\% (116,481)       & 46.9\% (112,522)         & 2.7\% (6,427)        & 1.9\% (4,499)        & 100\% (239,929)       \\ \bottomrule
\end{tabular}
\end{table}

\newpage
\section{RQ2 Topics and RQ3 Tech Related Risks Mapping}

\begin{table}[htpb]
\footnotesize
  \caption{Mapping of Technology Related Risks (RQ3) to Topics from Topic Modeling (RQ2)}
  \label{tab:topics}
  \centering
  \begin{tabular}{l|l}
    \hline
    \textbf{Technology Related Risks} & \textbf{Topics} \\
    \hline
    Sharing of Self-harm Thought and Triggering Contents & Self-harm\\ \hline
    Seeking Social Connection Led to Meeting Strangers Online& Social Connection \\ \hline
   Internet Facilitated Cyberabuse by Family and Others & Family Struggles\\ \hline
   Internet Facilitated Illegal and Risky behavior & Risk Behavior\\ \hline
    \hline
  \end{tabular}
  \end{table}

\end{document}